\documentclass[a4paper,11pt]{article}
\pdfoutput=1
\usepackage{jheppub}
\usepackage{calligra}
\usepackage{mathrsfs}
\usepackage{slashed}
\usepackage{graphicx}   
\usepackage{color}

\usepackage{feynmp}  
\usepackage{subcaption}
\newcommand{\CS}{\text{CS}}
\newcommand{\AdS}{\text{AdS}}
\newcommand{\dS}{\text{dS}}
\newcommand{\BMS}{\text{BMS}}
\newcommand{\XBMS}{\text{XBMS}}

\newcommand{\CFT}{\text{CFT}}
\newcommand{\Mink}{\text{Mink}}
\newcommand{\Mp}{\text{M}^{\text{Pl}}}
\newcommand{\Poincare}{\text{Poincare}}
\newcommand{\gl}{\text{global}}
\newcommand{\GR}{\text{GR}}
\newcommand{\CGR}{\text{CGR}}
\newcommand{\eff}{\text{eff}}
\newcommand{\Virasoro}{\text{Vir}}
\newcommand{\VirSquared}{\Virasoro_+ \times \Virasoro_-}
\newcommand{\weq}{\underset{\text{Weyl}}{\sim}}

\title{\center Asymptotic Symmetries, Holography \\
and Topological Hair}

\author{~ Rashmish K. Mishra and Raman Sundrum}
\affiliation{Maryland Center for Fundamental Physics, Department of Physics \\
 University of Maryland, College Park, MD 20742.}
\emailAdd{rashmish@umd.edu}
\emailAdd{raman@umd.edu}

\abstract
{
Asymptotic symmetries of AdS$_4$ quantum gravity and gauge theory are derived by coupling the holographically dual CFT$_3$  to Chern-Simons gauge theory and 3D gravity in a ``probe'' (large-level) limit. Despite the fact that the three-dimensional AdS$_4$ boundary as a whole is consistent with only finite-dimensional asymptotic symmetries, given by AdS isometries, infinite-dimensional symmetries  are shown to arise in circumstances where one is restricted to boundary subspaces with effectively two-dimensional geometry. A canonical example of such a restriction occurs within the 4D subregion described by a  Wheeler-DeWitt wavefunctional of AdS$_4$ quantum gravity. An AdS$_4$ analog of Minkowski ``super-rotation'' asymptotic symmetry is probed by 3D Einstein gravity, yielding CFT$_2$ structure (in a large central charge limit), via $\AdS_3$ foliation of $\AdS_4$ and the  $\AdS_3/\CFT_2$ correspondence. The maximal asymptotic symmetry is however probed by 3D {\it conformal} gravity. Both 3D gravities have Chern-Simons formulation, manifesting their topological character. Chern-Simons structure is also shown to be emergent in the Poincare patch of AdS$_4$, as soft/boundary limits of 4D gauge theory, rather than ``put in by hand'' as an external probe. This results in a finite effective Chern-Simons level. Several of the considerations of asymptotic symmetry structure are found to be simpler for AdS$_4$ than for  Mink$_4$, such as  non-zero 4D particle masses, 4D non-perturbative ``hard" effects, and consistency with unitarity. The last of these in particular is greatly simplified because in some set-ups the time dimension is explicitly shared by each level of description: Lorentzian AdS$_4$, CFT$_3$ and CFT$_2$.  Relatedly, the CFT$_2$ structure clarifies the sense in which the infinite asymptotic charges constitute a useful form of ``hair'' for black holes and other complex 4D states. An AdS$_4$ analog of Minkowski ``memory'' effects is derived, but with late-time memory of earlier events being replaced by (holographic) ``shadow'' effects.  Lessons from AdS$_4$ provide hints for better understanding Minkowski asymptotic symmetries, the 3D structure of its soft limits, and Minkowski holography. 
}

\begin{document}
\begin{flushright}
UMD-PP-017-22 \\
\end{flushright}
\maketitle
\flushbottom

\section{Introduction}
\label{sec:intro}

In gravitational and gauge theories, Asymptotic Symmetries (AS) are diffeomorphisms and gauge transformations that preserve the asymptotic structure of spacetime while still acting non-trivially on asymptotic dynamical data. They include isometries of spacetime and the standard global charges arising from gauge theory, but they can be larger. Famously, 4D Minkowksi spacetime (Mink$_4$) has an infinite-dimensional spacetime AS algebra (see~\cite{Strominger:2017zoo} for a recent review). This was originally identified as the BMS algebra of super-translations~\cite{Bondi:1962px,Sachs:1962wk}, but has been extended more recently to include super-rotations as a subalgebra~\cite{Barnich:2009se}. We refer to this extended algebra in 4D as XBMS$_4$. The ongoing challenge since discovery of these symmetries has been to understand their physical significance and utility. 

Considerable progress has been made in this regard by the discovery that the associated large diffeomorphisms and gauge transformations arise as soft limits of physical gravitational and gauge fields emerging from scattering processes~\cite{Strominger:2013lka,He:2014cra,He:2015zea,Kapec:2015ena,Strominger:2015bla,Strominger:2013jfa,He:2014laa,Kapec:2014opa,Lysov:2014csa,Mohd:2014oja,Dumitrescu:2015fej}, as captured by the Weinberg Soft Theorems~\cite{Weinberg:1965nx,Low:1958sn,Burnett:1967km,White:2011yy,Cachazo:2014fwa}. The infinite-dimensional AS then describe the soft field dressing of a hard process, and are sensitive to the passage of charge/energy-momentum as a function of angle, through ``memory'' effects~\cite{Bieri:2013hqa,Pasterski:2015zua,Susskind:2015hpa,Zeldovich:1974abc,Braginsky:1987abc,Christodoulou:1991cr,Strominger:2014pwa,Pasterski:2015tva,Zhang:2017geq,Zhang:2017rno}. 
This generalization of the usual overall charge/energy-momentum conservation laws has led to the suggestion that AS charges can act as a new subtle form of ``hair'' that can characterize black holes (or other complex states), giving a finer understanding of black hole entropy and information puzzles~\cite{Hawking:2016msc,Hawking:2016sgy,Carney:2017jut,Strominger:2017aeh}. 

The fact that the super-rotation subalgebra of $\Mink_4$ AS has a ${\rm Virasoro} \times \overline{{\rm Virasoro}} $ form (${\rm Vir} \times \overline{\rm Vir}$), while gauge theory gives rise to Kac-Moody  (KM) subalgebras, is highly reminiscent of Euclidean two-dimensional conformal field theories (ECFT$_2$)~\cite{Barnich:2009se}. Indeed such a ECFT$_2$-like structure living on the celestial sphere was discovered~\cite{Kapec:2016jld,Cheung:2016iub}, AS charges arising as Laurent expansion coefficients of a 2D holomorphic stress tensor and other currents. A straightforward derivation~\cite{Cheung:2016iub} follows by foliating Mink$_4$ by 3D de Sitter spacetimes (dS$_3$) and hyperbolic spaces~\cite{deBoer:2003vf,Solodukhin:2004gs,Chien:2011wz,Campiglia:2015qka,Costa:2012fm}, more suggestively considered as the Euclidean continuation of 3D anti-de Sitter (EAdS$_3$). 4D fields can then be ``Kaluza-Klein'' (KK) reduced by separation of variables into 3D (EA)dS$_3$ fields, with a continuum of 3D masses, $m^{\text{KK}}_3 > 0$. In this language, 4D S-matrix elements map to boundary (EA)dS$_3$ correlators~\cite{deBoer:2003vf}, the associated 4D LSZ-reduced Feynman diagrams mapping to 3D Witten diagrams (modulo superpositions). Most importantly, the 3D massless limit, $m_3 \rightarrow 0$, corresponds to 4D soft limits, in particular the soft limit of 4D gauge theory yielding 3D Chern-Simons (CS) gauge fields, and the subleading soft limit of 4D General Relativity (GR$_4$) fields yielding GR$_3$ (which also has a CS formulation~\cite{Witten:1988hc}) on (EA)dS$_3$.  The basic grammar of (EA)dS$_3$/ECFT$_2$~\cite{Strominger:2001pn} then yields the ECFT$_2$-like structure. The 3D CS fields ``live'' on the boundary of 4D spacetime.

~

Despite these recent developments, several important questions and puzzles remain:

$ \bullet$ A central question is how fully the axioms of CFT$_2$ are realized in the structure underlying AS. In particular, it has not been clear what the values of the associated central charge and KM levels are, whether zero, infinite or finite. This question is not answerable at the AS level of discussion which focuses on {\it external} CS/soft fields, since the central charge and levels are probed by internal CS/soft lines (at tree level). It was argued in Ref.~\cite{Cheung:2016iub}, that a central charge would be IR sensitive to the experimental delineation between ``soft'' and ``hard'', but this was not fully clarified.

$ \bullet$ The ECFT$_2$ structure is not consistent with being the Euclidean continuation of a unitary CFT$_2$, much as in the dS/ECFT context. It is an open question as to how the unitarity of the Mink$_4$ quantum gravity (QG) S-matrix is encoded in the ECFT$_2$ correlators.

$ \bullet$  The subleading soft limit of GR$_4$ leads to the super-rotation subalgebra of Mink$_4$ AS, and is elegantly encoded in GR$_3$, which has a $SO(3,1)$ CS formulation, but the leading soft limit and the associated super-translations do not have a CS formulation~\cite{Cheung:2016iub}. Naively, the $SO(3,1)$ Lorentz gauge group should be extended to the full Poincare group $ISO(3,1)$ as the CS gauge group in order to include (super-)translations, but 
$ISO(3,1)$  lacks the requisite quadratic invariant to construct a CS action. Relatedly, Ref.~\cite{Cheung:2016iub} found that the ECFT$_2$ current, 
whose Laurent expansion yields super-translations, is non-primary. Therefore there is an open question as to what the 3D characterization of subleading and leading soft GR$_4$ fields is that leads to XBMS$_4$ in a unified way.

$ \bullet$  Previous discussions of memory effects describe them in classical terms, while the hallmark of CS theories are quantum mechanical topological effects that generalize the Aharonov-Bohm effect~\cite{Aharonov:1959fk,Moore:1991ks,Wen:1990iu}. These two views of memories need to be better reconciled.

$ \bullet$ The connection of AS to 3D CS characterization of soft fields hints at  a possible connection to a 3D holographic duality with Mink$_4$ QG, but this connection has not been spelled out.

$ \bullet$  It is very attractive to contemplate AS charges as a new rich form of ``hair'' for black holes or other 4D states. But such a role is still unclear, and being debated~\cite{Bousso:2017dny,Donnelly:2017jcd}. 

~

In this paper, we make some progress on all these fronts within a more transparent context, 
by generalizing the notion of AS to 
AdS$_4$ QG and gauge theory. Primarily this is because we know the 3D holographic dual of AdS$_4$ is CFT$_3$~\cite{Maldacena:1997re,Gubser:1998bc,Witten:1998qj,Aharony:1999ti,Polchinski:2010hw,Sundrum:2011ic,Penedones:2016voo}, and there is a natural way to connect this to CS and GR$_3$, and from this to CFT$_2$ and infinite-dimensional AS. Yet by standard analysis the AS of asymptotically 
AdS$_4$ GR only consist of the finite-dimensional isometries~\cite{Ashtekar:1999jx}, $SO(3,2)$, in sharp contrast to the infinite-dimensional AS of asymptotically Mink$_4$ GR. Let us sketch why this is the case. 

First consider $\text{Mink}_4$, 
\begin{align}
ds^2_{\Mink_4} &= 
\frac{1}{\cos^2 u_+ \: \cos^2 u_-} 
\left(du_+ \: du_- - \frac{1}{4} \sin^2 (u_+ - u_-) \: d \Omega^2_2\right),
\qquad u_\pm = \tan^{-1}(t\pm r),
\end{align}
where $d \Omega^2$ is the usual metric of the angular sphere.
We see that at the boundary of $\text{Mink}_4$, $u_+ = \pm \pi/2$ and $u_- = \pm \pi/2$,
\begin{align}
ds^2_{\partial\text{Mink}_4} \:\: \weq \:\:     d\Omega^2_2,
\end{align}
where $\weq$  refers to Weyl equivalence, modulo which the notion of conformal boundary is defined. 
While the boundary manifold is three-dimensional, because of the null direction the geometry degenerates to being effectively two-dimensional.  A necessary condition for large diffeomorphisms to correspond to AS is that they
 preserve this boundary structure. In particular, these diffeomorphisms  include those reducing to {\it conformal isometries} on the boundary geometry, namely the infinite-dimensional conformal symmetries of the 
 2D angular sphere,  and correspond to the super-rotations.  
But in $\text{AdS}_4^{\gl}$, 
\begin{align}
ds^2_{\text{AdS}_4} = \frac{1}{\cos^2 \psi} \left( d\tau^2 - d\psi^2 - \sin^2 \psi \:  d\Omega_2^2 \right),
\end{align}
the boundary at $\psi = \pi/2$ has a fully three-dimensional geometry 
\begin{align}
ds^2_{\partial{\text{AdS}}_4^{\gl}} \weq d\tau^2 - d\Omega^2_2.  
\end{align}
The conformal isometries of this boundary $S^2 \times \mathbb{R}$, and hence AS of AdS$_4$, are just finite-dimensional $SO(3,2)$. By contrast, in the case of AdS$_3$, $\partial \AdS_3$ is obviously two-dimensional, famously with infinite-dimensional conformal isometries and AS~\cite{Brown:1986nw}.

\begin{figure}[h!]
\centering
\includegraphics[width=\textwidth]{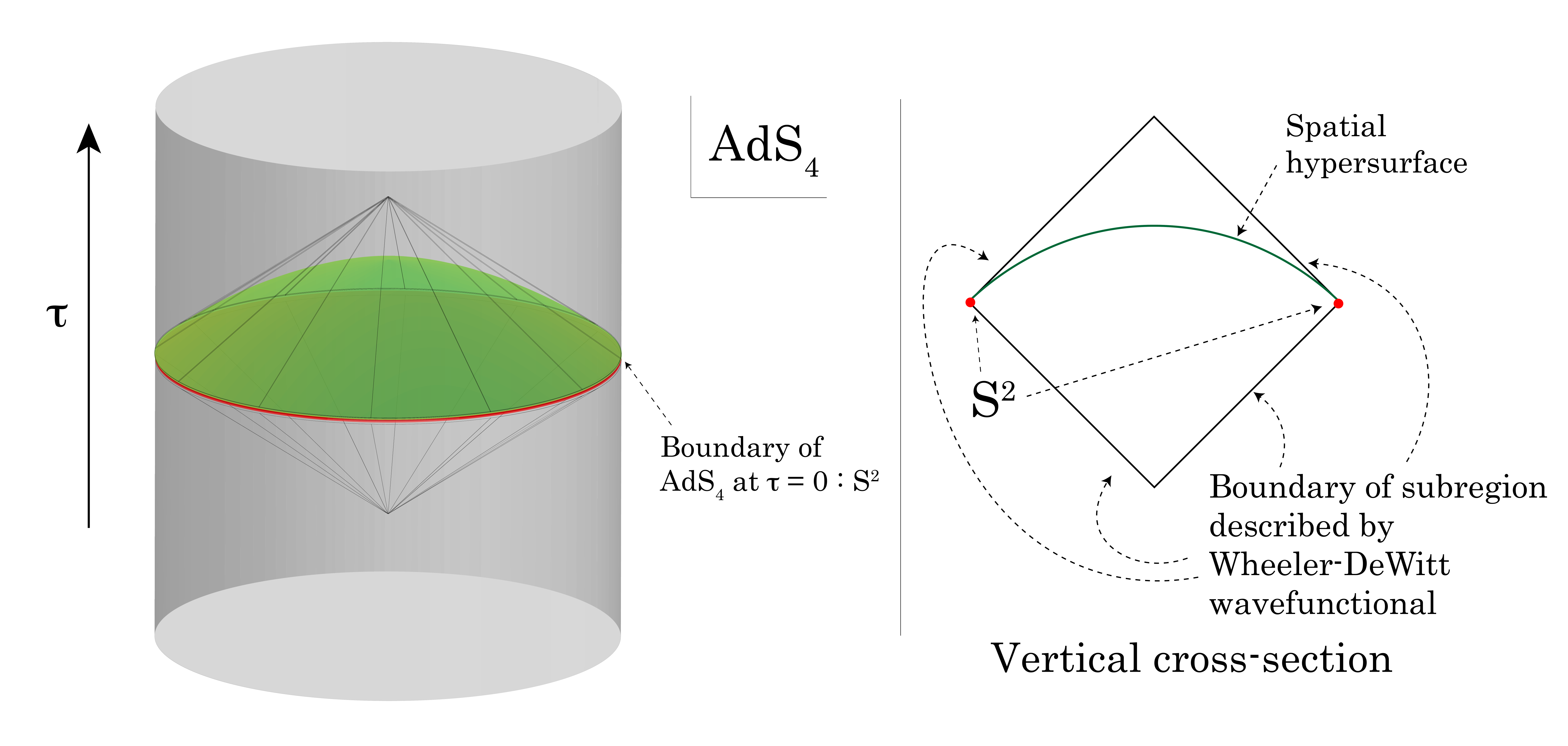}
\caption{$\CFT_3$ state living on $S^2$ at $\tau=0$ on $\partial \AdS_4$ (shown in red), dual to Wheeler-DeWitt wavefunctional describing the subregion of $\AdS_4$ enclosed by the black cones. This subregion is spanned by all spacelike hypersurfaces ending on this boundary $S^2$. An example of such a hypersurface is shown in green. A vertical cross-section is shown on the right.}
\label{fig:Wheeler-DeWitt-AdS4}
\end{figure}

Nevertheless,  there is a loop-hole to this no-go argument for infinite-dimensional  AdS$_4$ AS  if one is restricted to subspaces of $\partial \AdS_4$ with {\it two-dimensional} geometry, which we will see can happen for different physical reasons. Most straightforwardly, this is illustrated by the subregion of $\AdS_4$ described by a Wheeler-DeWitt QG wavefunctional, holographically dual to a quantum state of CFT$_3$ at some fixed time, as depicted in Fig.~\ref{fig:Wheeler-DeWitt-AdS4}. Its 3D boundary resembles the null boundary of Mink$_4$,  with effectively 2-dimensional geometry, reflecting the two-dimensional holographic geometry of $\partial \AdS_4$ at $\tau = 0$. This has infinite-dimensional conformal 
isometries, leading to infinite-dimensional AS.

~

The basic strategy of this paper will be to study CS gauge theory and GR$_3$ coupled to $\CFT_3$, where the $\CFT_3$ is (in isolation) the holographic dual of $\AdS_4$ QG, on a variety of 3D spacetimes ${\cal M}_3$:
\begin{equation}
S = S_{\text{CS}} + S_{\GR_3} + S_{\CFT_3} + \text{UV-completion}.
\end{equation} 
The $\CFT_3$ global internal symmetries are gauged by the CS sector,  and the  $\CFT_3$ spacetime symmetries are gauged by $\GR_3$. Such $\GR_3$ and CS $+$ matter theories are well-known to have infinite-dimensional AS~\cite{Brown:1986nw,Ashtekar:1996cd,Barnich:2006av,Spradlin:2001pw,Anninos:2010zf,Witten:1988hf,Elitzur:1989nr,Witten:1991mm,Gukov:2004id}. 
 In particular, when ${\cal M}_3 = \AdS_3$,  AdS$_3$/CFT$_2$ implies this setup is dual to CFT$_2$, where there is a standard connection of the 2D chiral currents and stress-tensor with infinite-dimensional KM and  $\VirSquared$ symmetries (briefly reviewed in Section~\ref{sec:CS-review}). The infinity of (AS) charges of CFT$_2$ (AdS$_3$) 
form a well-known type of 2D (3D) ``hair", operating on and finely diagnosing quantum states, 
in a manner generalizing  the action of ordinary conserved global charges. But now the $\CFT_3/\AdS_4$ duality of the 3D matter ``lifts" the AS charges and their utility to 4D.

 This construction yields three layers of description of the dynamics. The quarks and gluons of some 
 large-$N_{\text{color}}$ formulation of $\text{CFT}_3$ will be called for brevity, ``quarks''. The dual $\text{AdS}_4$ gravitons and matter are the ``hadrons'', composites of the 3D ``quarks'', the 4D fields being equivalent to KK towers of 3D ``hadronic'' states related by 3D conformal symmetry. 
  The well-defined $\partial$AdS$_3$ correlators will involve external  lines of these ``hadrons'', rather than ``quarks'' (as discussed in Section~\ref{sec:CS-plus-CFT3}). This is in complete analogy to the well-defined nature of hadronic S-matrix elements in Minkowski spacetimes, as compared with the provisional nature of the quark/gluon S-matrix.
 Even more fundamentally, the $\CFT_3$ ``quarks'' and the CS $+$ $\GR_3$ fields themselves are composites of the $\text{CFT}_2$ degrees of freedom, which we call  ``preons''. 
 AS charges are simple moments of these local ``preon'' degrees of freedom. 
  A nice feature here is that 
 time persists  at each layer of description, and hence unitarity is manifest at each stage. The 4D loop expansion (controlled by the expansion parameter $1/N_{\text{color}}$ in 3D) can be done to all orders without spoiling these results. Including 4D massive particles is straightforward, captured automatically by the CFT$_3$ description.

  We will show that even in the large-level limit, in which the CS and $\GR_3$ fields are decoupled, these AS remain as subtle charges of the matter sector, $\CFT_3$ (see Section~\ref{sec:LargeKappa}). 
  Because the $\CFT_3$ on $\AdS_3$ is dual to (half of) $\AdS_4$, 
    the 3D AS are inherited as AS of $\AdS_4$ QG and gauge theory. 
   From the 4D perspective (Section~\ref{sec:AdS4-AdS3-Corr}), the ``hadronic'' $\partial \AdS_3$ correlators which manifest the infinite-dimensional AS are also  $\partial \AdS_4$ correlators, but not of the standard form. 
    In particular, the $\partial \AdS_3$ endpoints are restricted to a submanifold of $\partial \AdS_4$ with two-dimensional geometry, one natural realization of the loop-hole mentioned earlier in the no-go argument for infinite-dimensional AdS$_4$ AS (Section~\ref{sec:Evading-NoGo}).

The AS of AdS$_4$ are in fact closely analogous to those of $\text{Mink}_4$, in particular the $\VirSquared$ can be viewed as analogous to $\text{Mink}_4$ super-rotations. The analog of $\text{Mink}_4$ super-translations is subtler. We will show (Section~\ref{sec:CGR3}) that these can be incorporated by replacing GR$_3$ by 3D \textit{conformal} gravity ($\text{CGR}_3$)~\cite{Deser:1982vy,Deser:1981wh}, which also has a $SO(3,2)$ CS formulation~\cite{Witten:1988hc,Horne:1988jf}. In the case of ${\cal M}_3 = \AdS_3$, this leads to an extension of the AS by a KM algebra~\cite{Afshar:2011qw}. 
But the full AS of $\AdS_4$  is even larger, because CFT$_3$ on $\AdS_3$ only projects half of $\AdS_4$. 
The technically simplest approach to the full AS structure is taken by switching to ${\cal M}_3 = \Mink_3$, where the 
dual of the CFT$_3$ is given by the Poincare patch, $\AdS_4^{\Poincare}$ (Section~\ref{sec:PoincarePatch}). While not the entirety of 
$\AdS_4^{\text{global}}$, it shares all of its (infinitesimal) isometries, and hence exhibits the full AS algebra. 
This full AS structure allows us to run the connection to 
 $\AdS_4/\CFT_3$ duality in reverse: if one begins by identifying the AS of AdS$_4$ in  CGR$_3$ ($SO(3,2)$ CS) form, the only form of compatible matter that can couple to CGR$_3$, respecting its Weyl invariance, is CFT$_3$.  In this sense, the holographic grammar follows from the AS structure.

The Poincare patch provides other simplifications. It
gives the most straightforward 4D dual picture when GR$_3$ is {\it not yet decoupled} from 
CFT$_3$, namely a lower-dimensional Randall-Sundrum 2 (RS2) construction~\cite{Randall:1999vf}, with a 3D ``Planck brane'' in a 4D bulk~\cite{Emparan:1999wa}. The GR$_3$ then incarnates as the localized gravity of RS2. 
The familiarity of RS2 helps to make an important contrast. We have argued above, and in the body of this paper, that the infinite-dimensional AS are most readily recognized as coming from 3D GR$_3$/CS fields, and yet are interesting 
 because we can ``lift" them beyond three dimensions.  But there appears to be an even easier way to arrange this, by just considering gravitational theories in higher-dimensional product spacetimes of the form $\Mink_3 \times X$ or 
$\AdS_3 \times X$, where $X$ is some compact manifold. Under Kaluza-Klein reduction to $\Mink_3$ or $\AdS_3$, 
such theories would have a GR$_3$ 3D-massless mode, which would again yield infinite-dimensional symmetries. The distinction with what we are doing here is that such product theories would not have a non-trivial decoupling limit for the GR$_3$ fields. That is, we cannot sensibly remove the GR$_3$ subsector in some limit while keeping the rest of the physics fixed. But RS2 with 4D bulk is dual to GR$_3 + \CFT_3$, and there is a limit in which the 3D gravitational coupling vanishes, leaving a fixed limiting $\CFT_3$, dual to $\AdS_4^{\Poincare}$ QG. In other words, we will argue that GR$_3$/CS has a tight connection with AS structure on the one hand and with 3D holography of the 4D QG on the other. But this only takes place in higher-dimensional theories where the GR$_3$ subsector has a decoupling limit. Higher-dimensional product spacetimes are not of this type.

 The Poincare patch also provides the stage to simply derive the {\it emergence} of 
 CS gauge fields as helicity-cut soft/boundary limits of AdS$_4$ gauge fields (CFT$_3$ composites), which couple to charged modes (Section~\ref{sec:Emergent}). 
 In this way, the CS structure is not put in ``by hand'' and then removed by a large-level limit, but rather describes
 a subsector
   of the pure CFT$_3$/AdS$_4$, with a finite but subtle type of CS level.  We will see that the effective CS gauge fields mediate  analogs of the ``memory" effects identified in Mink$_4$,  which we call ``shadow" effects since their relationship to the holographically emergent spatial direction is analogous to the relationship of memory effects to time. 
 
 For the CFT$_3$ to project all of AdS$_4^{\text{global}}$, we must choose ${\cal M}_3 = S^2 \times \mathbb{R}$ (Section~\ref{sec:s2-times-R-cutInTime}), but this closed universe does not have an asymptotic region or boundary to straightforwardly display AS. The AS arise by cutting at some point in time (say zero), so that the wavefunctional is given by functional integration up to that point, that is on ${\cal M}_3 = S^2 \times \mathbb{R}^{-}$ (where the last factor refers to only negative values of time). This yields precisely the holographic dual of the Wheeler-DeWitt wavefunctional in $\AdS_4$, briefly discussed above.

Finally, it is obviously of interest to ask how to translate the insights of AdS$_4$ AS back to Mink$_4$ (Section~\ref{sec:futureDirections}).
  A strategy is suggested by the argument of Section~\ref{sec:PoincarePatch} for deriving the holographic grammar of AdS$_4$ from its AS structure. In Mink$_4$ we are ignorant of the former but know the latter, so the analogous steps should yield new insight into Mink$_4$ holography. The first step is to give the 3D characterization of the full AS and soft fields of Mink$_4$ QG, in analogy to identifying CGR$_3$ for AdS$_4$. Currently this is not known for the 
   Mink$_4$ super-translations, although super-rotations take a simple GR$_3$ form. We will provide some concrete guesses as to how to obtain the full 3D structure, which will then form the ``mold" for a compatible holographic form of (hard) matter.

\section{Lightning Review of CS/GR$_3$ AS and $\CFT_2$ Currents}

\label{sec:CS-review}

CS theories, including GR$_3$ in CS form, are famously gauge invariant and topological, insensitive to the geometry on 3D spacetime ${\cal M}_3$, except at the boundary $\partial {\cal M}_3$ where local degrees of freedom emerge, exhibiting infinite-dimensional AS. We briefly review how this happens for ${\cal M}_3 = \AdS_3$, where the boundary structure and AS are just those of the dual $\CFT_2$. 
Concretely, we write the metric in the form
\begin{align}
ds^2_{\text{AdS}_3^{\gl}} 
=
\frac{d\tau^2 - d\rho^2 - \cos^2\rho \:d\phi^2}{\sin^2 \rho}, \qquad R_{\text{AdS}_3} \equiv 1.
\end{align}
A point $y^\mu$ in $\text{AdS}_3$ is represented by the coordinates $(\tau, \phi, \rho)$, where $-\infty < \tau < \infty$, $0 \leq \phi < 2\pi$ and $0 < \rho \leq \pi/2$. The space of $\text{AdS}_3$ is conformally equivalent to $S^2/2 \times \mathbb{R}$, and the boundary $\partial\text{AdS}_3$ is at $\rho=0$ in these coordinates.

\subsection{Non-abelian CS gauge theory}

We begin with internal CS gauge theory, 
\begin{align}
S_{\text{CS}}
=
\frac{\kappa}{4\pi}
\int d^3y \: \epsilon^{\mu\nu\rho} \:\text{Tr} 
\: \left(A_\mu \partial_\nu A_\rho  + \frac23 A_\mu A_\nu A_\rho\right)\:,
\end{align}
where $A_\mu \equiv A^a_\mu\:t^a$, $t^a$ are the generators of the CS gauge group, $\text{Tr}(t^a t^b) = \delta^{ab}$, and  $\kappa$ is the CS level.
\\
This action is metric-independent and gauge-invariant in the $\AdS_3$ ``bulk'', but since gauge-invariance depends on integration by parts it is violated on the boundary, $\partial \AdS_3$. This implies that ``gauge orbit'' degrees of freedom ``live'' on this 2D boundary, $\rho = 0$, which is the root of the equivalence of the CS gauge sector to a 2D Wess-Zumino-Witten (WZW) current-algebra sector on the boundary~\cite{Witten:1988hf,Elitzur:1989nr,Witten:1991mm,Gukov:2004id}. 

It is convenient to use light-cone coordinates in the boundary directions, 
\begin{equation}
z^\pm \equiv \tau \pm \phi.
\end{equation}
The equations of motion read
\begin{align}
\delta S_{\text{CS}} 
&=
\frac{\kappa}{4\pi}\:  {\rm Tr}  \int d^3 y \epsilon^{\mu \nu \rho} \left(\delta A_{\mu} F_{\nu \rho}\right) 
\nonumber \\
&+  \frac{\kappa}{2\pi}{\rm Tr} \int dz ^+ \: dz^- \: \:
\Big( 
\delta A_-(\rho=0)  \: A_+(\rho=0) - \delta A_+(\rho=0) \: A_-(\rho=0) 
\Big) = 0\:.
\label{eq:boundaryVariationCS}
\end{align}
This implies boundary conditions,  $A_{\pm}(\rho = 0) = 0$. Further, bulk gauge invariance can be used to go to  the axial gauge: $A_\rho = 0$. With this the boundary conditions are too stringent, giving   $A_\mu = 0$ 
throughout $AdS_3$ as the only solution to the first order equations.

We can modify the boundary conditions to constrain just one linear combination of boundary components of $A$, say $A_-$. To accomplish this we can add a boundary term to the action,
 \begin{align}
S_{\partial \AdS_3} 
=
-\frac{\kappa}{2\pi} \:\int dz^+ \: dz^- \: \text{Tr}\:
\Big(A_-(\rho=0) \:A_+(\rho = 0) \Big)\:.
\end{align}
(While this explicitly violates gauge invariance, recall  the bulk action is already not gauge-invariant on the boundary.)
In the presence of this term, the total boundary contribution to the variation of the action is given by 
\begin{align}
\delta S_{\text{total}}\bigg\rvert_{\partial\AdS_3}
=
-\frac{\kappa}{\pi}\: \int dz^+ \: dz^- \: \text{Tr}\:
\Big( \delta A_+ \: A_- \Big),
\end{align}
implying the boundary condition $A_-(z^+, z^-, \rho = 0) =0$.  But now $A_+(z^+, z^-, \rho = 0)$ is unconstrained, consistent with non-trivial solutions (in the presence of matter).

Even though we have fixed axial gauge $A_{\rho} = 0$, we must retain the $A_{\rho}$ equation of motion, 
\begin{equation}
F_{+-} = 0,
\end{equation}
away from any matter sources, where $F$ is the non-abelian field strength. 
Evaluating this on the boundary, and using the boundary condition $A_- = 0$, 
\begin{align}
F_{+-} \underset{\rho \rightarrow 0}{\longrightarrow} \partial_- A_+ = 0.
\end{align}
The dual $\CFT_2$ current, 
\begin{equation}
j_+(z^+, z^-) = \underset{\rho \rightarrow 0}{\lim} A_+(z^+, z^-, \rho),
\end{equation}
is therefore chirally conserved,
\begin{align}
\partial_- j_+ = 0\:,  ~ j_+ = j_+(z^+).
\end{align}

The Fourier components define AS charges,
\begin{align}
j_+(z^+) = \sum_{n \in \mathbb{Z}} Q_n^{a+}(\tau)  t^a \: e^{i n \phi}\: \:,
\end{align}
which are angle-dependent ``harmonics'' of the conserved global charges, $Q^{a+}_0$.
The $\tau$ dependence of $Q^{a+}_n(\tau)$ follows by the fact that $j_+$ is a function of $z^+ = \tau + \phi$ only, simply given by
\begin{align}
Q^{a+}_n(\tau) \propto e^{i n \tau}\:.
\label{eq:KM-Charge-Tau-Dependence}
\end{align}
In the $Q_n^+$ basis, the simple structure of $j$ correlators within $\partial \AdS_3$ Witten diagrams takes the form of a Kac-Moody algebra (at $\tau =0$),
\begin{align}
\left[\: Q_n^{a+},\: Q_m^{b+} \: \right] &= \sum_c \: i \: f^{abc}\: Q^{c+}_{n+m} + \kappa \: n \: \delta^{ab} \: \delta_{n+m,0}
\end{align}
where $f^{abc}$ are the structure constants, and the CS level $\kappa$ sets the central extension. 
The non-abelian first term on the right-hand side reflects the non-abelian CS interaction, while the central extension second term on the right-hand side reflects the CS ``propagation''.

\subsection{$\GR_3$}

Consider next the case of 3D gravity on $\AdS_3$, which can be formulated as a $SO(2,2)$ CS theory in terms of dreibein and spin connection variables~\cite{Witten:1988hc}, with level 
\begin{equation}
\kappa_{\text{grav}} = \Mp_3\:R_{\text{AdS}_3}.
\end{equation}
The dreibein VEVs lock the six $SO(2,2)$ global generators $L^\pm_{n=-1,0,1}$ to the $\AdS_3$ isometries. The action of these generators at the boundary of $\AdS_3$ is given by
\begin{align}
L^\pm_n \underset{\partial \AdS_3}{\longrightarrow} e^{i n z^\pm}\:\partial_\pm\:, \qquad n = \pm 1, 0\:.
\end{align}
Analogous to the case of internal CS gauge symmetries, the 
stress tensor components are chiral, and their Fourier modes give angle-dependent ``harmonics'' of the above $SO(2,2)$ global symmetries, 
\begin{align}
t_{++}(z^+) = \sum_{n \in \mathbb{Z}} L_n^+(\tau) e^{i n \phi}\:,
\qquad t_{--}(z^-) = \sum_{n \in \mathbb{Z}} L_n^-(\tau) e^{-i n \phi}\:,
\end{align}
where the $\tau$ dependence of $L_n^+(\tau)$ is fixed:
\begin{align}
L_n^+(\tau) \propto e^{i n \tau}\:.
\end{align}
These AS charges now form a $\VirSquared$ algebra~\cite{Brown:1986nw} generalizing the $SO(2,2)$ isometries, as opposed to a KM algebra if there had been no VEVs (as reviewed in Ref.~\cite{Fitzpatrick:2016mtp}),  
\begin{align}
\left[L_m^\pm, L_n^\pm \right]
= \left(m-n\right)L^\pm_{m+n} + \frac{c}{12}\: m(m^2-1) \: \delta_{m+n, 0}\:, \qquad m,n ~ {\rm integer}.
\end{align}
The central charge is given by
\begin{equation}
c = \kappa_{\text{grav}} = \Mp_3 R_{\AdS_3}.
\end{equation}
Again, the two terms on the right-hand side are the 2D reflection of  the non-abelian interaction of GR$_3$ and the free ``propagation''.  
\\
The charges $L_n$ have a non-zero commutator with internal KM charges $Q_n$,
\begin{align}
\left[ L^+_m, Q_n^+ \right] = n Q^+_{m+n}
\end{align}
while the commutator between $+$ and $-$ charges vanishes. Given that $m=0$ measures the energy corresponding to $\tau$ translational symmetry: $E_\tau = L^+_0 + L^-_0$, it follows that
\begin{align}
\left[ L^+_0 + L^-_0, Q_n^+ \right] = n\: Q^+_{n},
\end{align}
matching our earlier observation that $Q^+_n \propto e^{i n \tau}$.

\section{Holographic Matter coupled to CS/$\GR_3$ on $\AdS_3$}

\label{sec:CS-plus-CFT3}

We now couple CS and GR$_3$ to 3D matter in the form of CFT$_3$, all living on asymptotic AdS$_3$. 
The $\CFT_3$ is chosen such that when living (in isolation) on 
 $\partial\AdS_4$ = $S^2 \times \mathbb{R}$ it is holographically dual to some $\AdS_4$ QG and gauge theory.

\subsection{$\CFT_3$ in isolation on AdS$_3$}

We begin by noting that 
\begin{equation} 
\AdS_3 \weq S^2/2 \times \mathbb{R}, 
\end{equation}
  where $\weq$ denotes Weyl equivalence, and $S^2/2$ denotes the hemisphere. Since 
  $\partial\AdS_4$ = $S^2 \times \mathbb{R}$  is only defined up to Weyl equivalence, this suggests that CFT$_3$ on 
  AdS$_3$ is holographically dual to {\it half} of AdS$_4$, as follows.

\begin{figure}
  \centering
  \includegraphics[width=.7\linewidth]{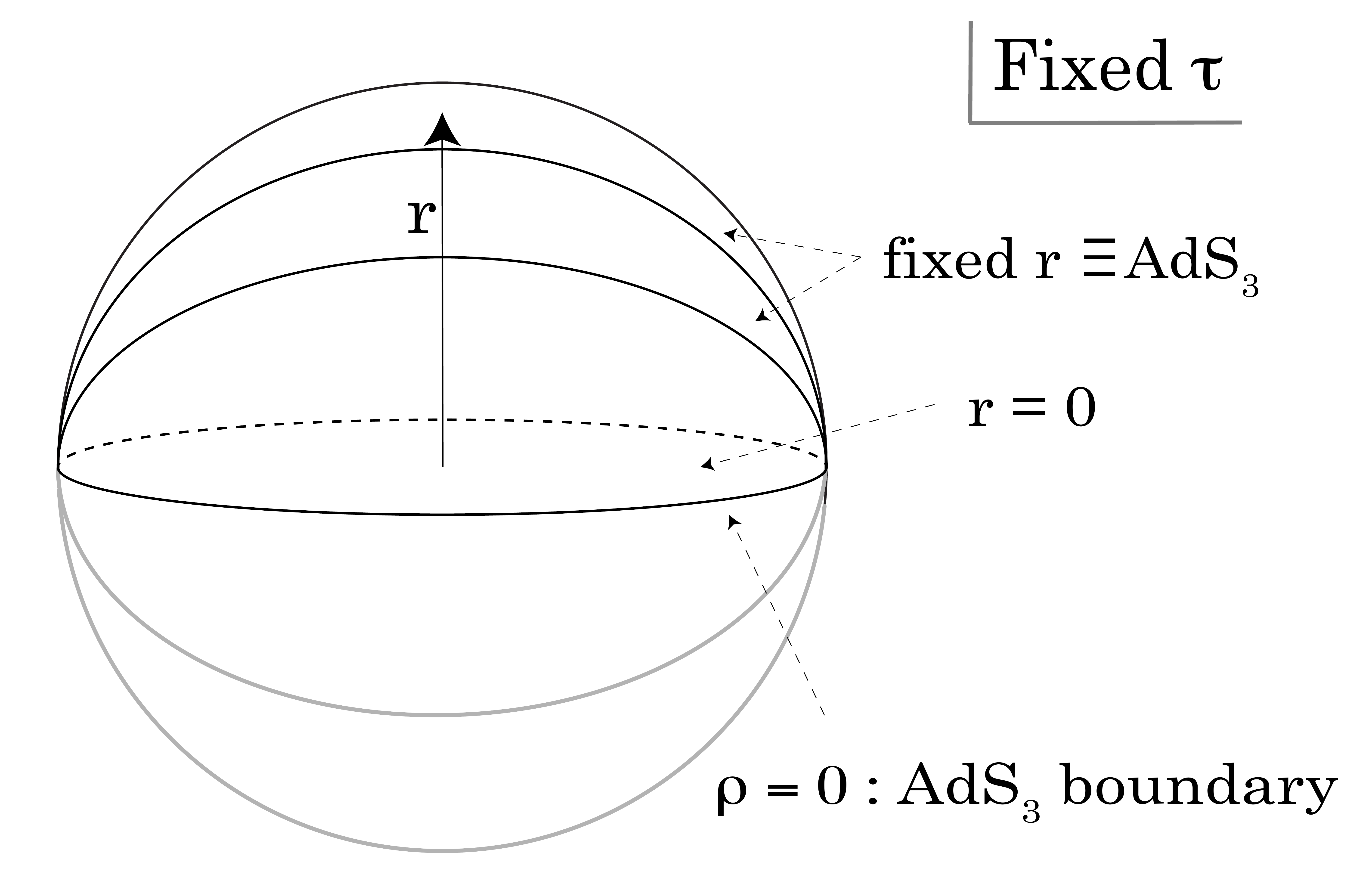}
  \caption{$\AdS_3$ foliation of $\AdS_4$ in global coordinates. A $\CFT_3$ on $\AdS_3$ projects only the upper half of $\AdS_4$.}
  \label{fig:AdS4-Global}
\end{figure}

It is useful to use $\AdS_4$ coordinates exhibiting an $\AdS_3$ foliation~\cite{Karch:2000ct},
\begin{align}
ds^2_{\AdS_4} &= -dr^2 + \cosh^2 r \: ds^2_{\AdS_3},\qquad r \in \mathbb{R},\qquad R_{\AdS_4} = 1
\nonumber
\\
ds^2_{\AdS_3} &= \frac{1}{\sin^2 \rho} \left( d\tau^2 - d\rho^2 - \cos^2 \rho \: d\phi^2 \right),\qquad R_{\AdS_3} = 1.
\end{align}
The AdS$_3$ coordinates $(\tau, \phi, \rho)$ have the ranges $ -\infty < \tau < \infty, 0 < \rho \leq \pi/2$, $0 \leq \phi < 2\pi$, while the fourth dimension coordinate $r$ takes all real values. Ref.~\cite{Bousso:2001cf} argued (translating their analysis down a dimension to the set-up of interest here) that CFT$_3$ states on AdS$_3$, reflecting off $\partial \AdS_3$ are dual to 
 AdS$_4$ particles in the region $r > 0$ reflecting off the  $r = 0$ surface.   The specific boundary condition at $r=0$ is determined by the whether or not the CFT$_3$ ground state on AdS$_3$ preserves or spontaneously breaks the CFT$_3$ global symmetry. We will consider the case where the global symmetry is preserved, in which case we must choose Neumann boundary condition at $r=0$.  
 We denote the region $r>0$, holographically projected by CFT$_3$,
 by ``$\AdS_4/2$''. 
In the original $\AdS_4$ global coordinates the $\AdS_3$ foliation by constant $r$ hypersurfaces is depicted in  Fig.~\ref{fig:AdS4-Global}, where the restriction to $\AdS_4$/2 corresponds to keeping only the northern half of the coordinate ball, $r=0$ being the equatorial disc. The $\CFT_3$ lives on the boundary of this region, the upper hemisphere.

\begin{figure}
  \centering
  \includegraphics[width=.7\linewidth]{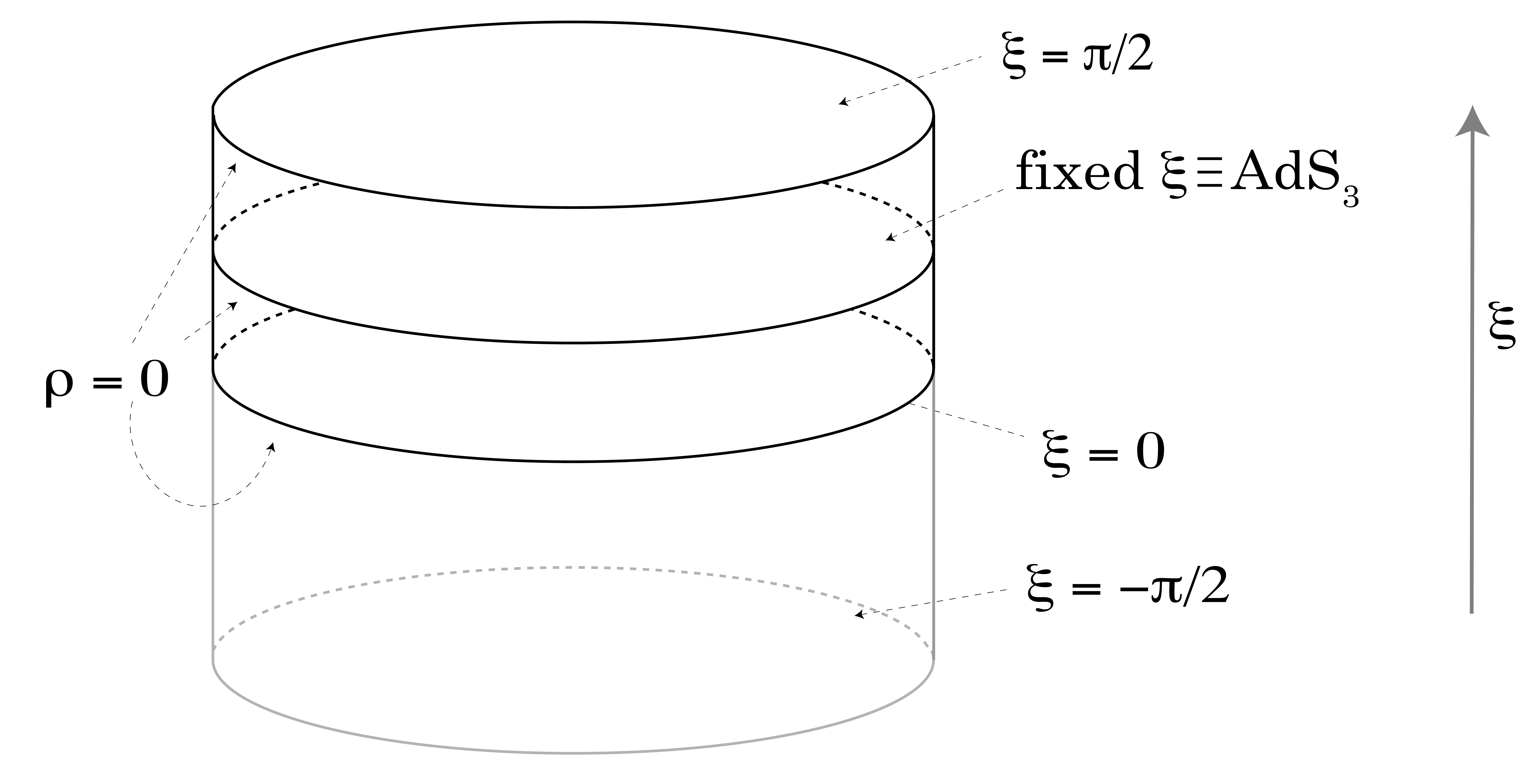}
  \caption{$\AdS_3$ foliation of $\AdS_4$ in ``product-space'' coordinates.}
  \label{fig:AdS4-TinCan}
\end{figure}

$\partial \AdS$ correlators are the classic diffeomorphism and gauge invariant observable in AdS QG,
just as the S-matrix is in Mink QG.  Here we are preparing to couple $\CFT_3$ on $\AdS_3$ to $\GR_3$ and CS, so we are interested in $\partial \AdS_3$ correlators. In this subsection however we are not yet including the gauging by $\GR_3$ and CS, focusing therefore on $\partial \AdS_3$  correlators of just the $\CFT_3$. 
In standard Minkowski QCD we have a provisional meaning for the S-matrix elements of quarks and gluons. But strictly speaking this is ill-defined because they are not asymptotic states. Instead we should more properly consider S-matrix elements of hadrons such as protons and pions. Similarly, with the $\CFT_3$, instead of ``quark'' $\partial \AdS_3$ correlators, we consider ``hadron'' $\partial \AdS_3$ correlators. Of course these ``hadrons'' are given precisely by the $\AdS_4$ dual. But now each $\AdS_4$ field contains many ``hadronic'' $\AdS_3$ mass eigenstates, which we can isolate by KK decomposition based on the $\AdS_3$ foliation. 

We illustrate this for the simple case of tree-level $\AdS_4$ Yang-Mills theory, with 4D field ${\cal A}$. For this purpose it is convenient to adopt what we call
 ``product-space'' coordinates. 
  Using the change of variables from $r$ to $\xi \equiv 2\tan^{-1}\left(\tanh (r/2)\right)$, the metric changes to
\begin{align}
ds_{\AdS_4}^2 &= f^2(\xi)\:\left(-d\xi^2 + ds_{\AdS_3}^2\right) \nonumber \\ 
\xi &= 2 \tan^{-1} \left( \tanh \frac{r}{2} \right), \qquad f(\xi) = \cosh r\:,
\label{eq:AdS4-xiCoordinates}
\end{align}
displaying Weyl-equivalence to the product geometry $\AdS_3 \times$ Interval.
The restricted region $\AdS_4$/2, $0 < r < \infty$ corresponds to $ 0 < \xi < \pi/2$ (in AdS units). Fig.~\ref{fig:AdS4-TinCan} shows this ``product-space'' representation of $\AdS_4$. 

\begin{figure}[h!]
\centering
\includegraphics[width=0.3\textwidth]{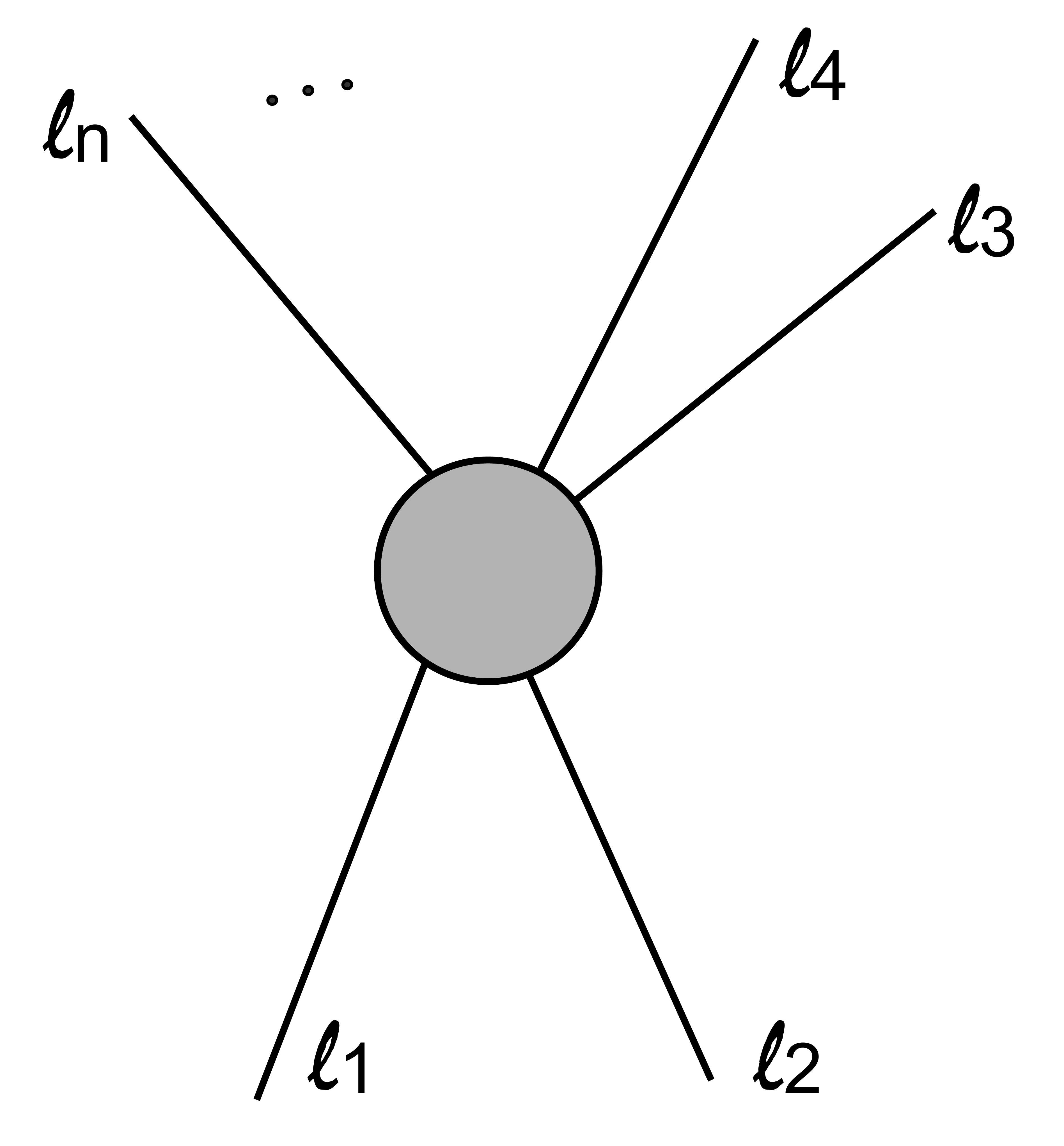}
\caption{Witten diagrams for $\partial \AdS_3$ correlators of KK modes ($\equiv \CFT_3$ ``hadrons''). External lines are $\AdS_3$ bulk-boundary propagators for masses proportional to $\ell$. Blob consists of $\AdS_3$ KK interactions and bulk-bulk lines (Fourier transformed in $\xi$ from $\AdS_4$).}
\label{fig:WittenDiagram}
\end{figure}

Because of the Weyl invariance of classical 4D Yang-Mills, the factor $f^2(\xi)$ is irrelevant and the spacetime is effectively  of product form. (Non-Weyl-invariant theories can also be KK-decomposed, but less straightforwardly.) In standard KK fashion, in axial gauge  $\mathcal{A}_\xi = 0$, the 4D Maxwellian field decomposes as
\begin{align}
\mathcal{A}_\mu(\tau, \phi, \rho, \xi) = \sum_{\ell \in \mathbb{Z}} A_\mu^\ell ( \tau, \phi, \rho) \cos ((2 \ell + 1 )\xi) \: ,
\end{align} 
where the $A_\mu^\ell ( \tau, \phi, \rho)$ are a tower of 3D Proca fields with $\AdS_3$ masses $2\ell + 1$ in units of AdS radius.
It is $\AdS_3$ Witten diagrams of these KK fields that correspond to ``hadron'' $\partial \AdS_3$ correlators.
 We depict such diagrams in 
 Fig.~\ref{fig:WittenDiagram}.
\subsection{CS and $\GR_3$ coupled to $\CFT_3$}
We now switch on $\text{CS}$ and $\text{GR}_3$, gauging any internal global symmetries of $\CFT_3$ (dual to 4D gauge symmetries) and the global spacetime symmetries and stress tensor of $\CFT_3$ (dual to 4D gravity),
so that $\partial \AdS_3$ correlators include the dual 2D chiral currents $j_\pm$, and stress tensor $t_{\pm\pm}$. The 3D KK modes discussed above are dual to local 2D primary operators $\mathcal{O}_{2D}^\ell$. All the $\CFT_2$ operators are composites of some 2D ``preon'' fields. 

\begin{figure}[h!]
\centering
\includegraphics[width=0.7\textwidth]{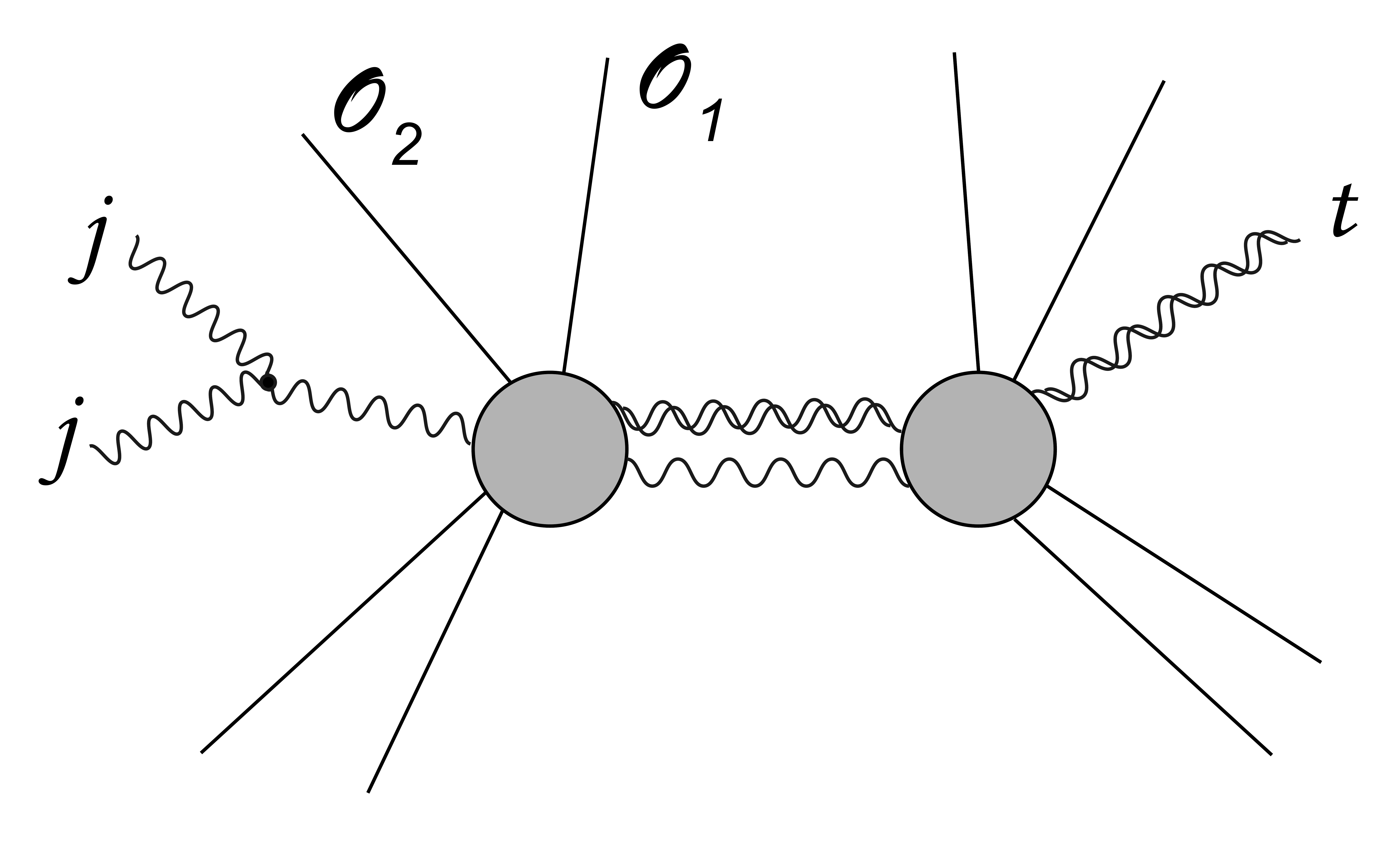}
\caption{Witten diagram for $\partial \AdS_3$ correlators, involving internal and external graviton/gluon lines.}
\label{fig:TypicalDiagram}
\end{figure}

A typical Witten diagram is shown in Fig.~\ref{fig:TypicalDiagram}, with CS and $\GR_3$ lines decorating the earlier purely ``hadronic'' diagrams.
Such 3D Witten diagrams
yield general $\CFT_2$ dual correlators, now including stress tensor and chiral currents,
 $\left< 0\right| T \{  j_\pm \dots t_{\pm \pm} \: \dots \mathcal{O}^\ell \dots \} \left| 0 \right>$. As we reviewed in Section~\ref{sec:CS-review}, this $\CFT_2$ has infinite-dimensional symmetries associated with its chiral currents and stress tensor.

\section{The Large-Level ``Probe'' Limit}

\label{sec:LargeKappa}

The decoupling of the CS and GR$_3$ sectors from $\CFT_3$ is accomplished by simply taking the large CS-level limit, $\kappa \rightarrow \infty$. (For decoupling the $\text{GR}_3$ sector this is equivalent to the large-central-charge limit of the Virasoro symmetry of the CFT$_2$ dual.)  
We will show that in this limit there is a remnant of the AS algebra that survives for  $\CFT_3$ alone, providing a new form of ``hair'' for the dual $\AdS_4$/2 states and black holes.

\subsection{Abelian CS}

\begin{figure}[h!]
\centering
\begin{subfigure}[b]{.48\textwidth}
  \centering
  \includegraphics[width=.9\linewidth]{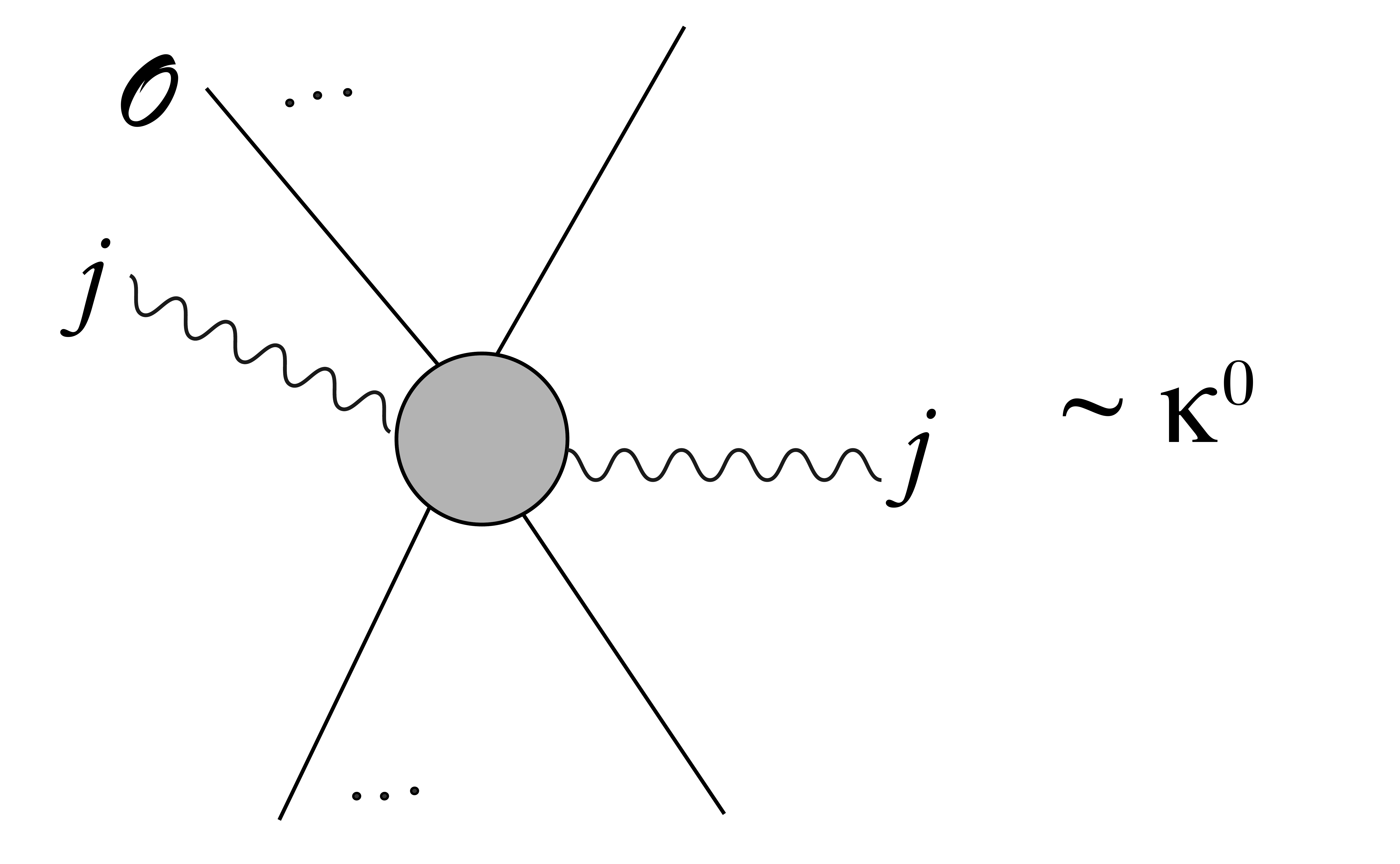}
  \caption{Leading Witten diagrams including external $\CFT_3$ lines.}
  \label{fig:SurvivingDiagram}
\end{subfigure}\hfill%
\begin{subfigure}[b]{.48\textwidth}
  \centering
  \includegraphics[width=.8\linewidth]{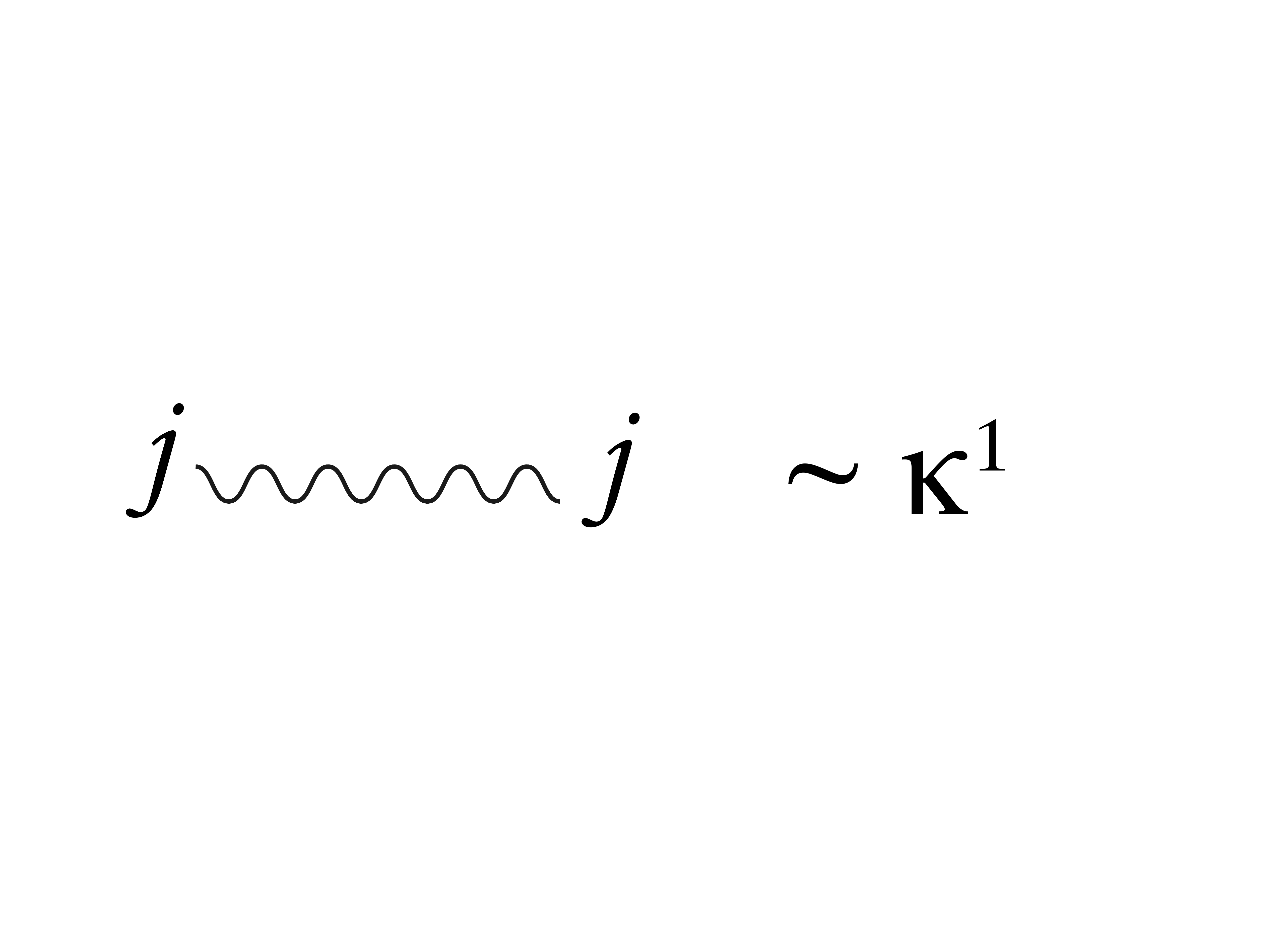}
  \caption{Leading Witten diagram without external $\CFT_3$ lines.}
  \label{fig:SpecialDiagram}
\end{subfigure}
\caption{Leading Witten diagrams in abelian CS theory for large $\kappa$.}
\label{fig:LargeKappaDiagrams-Abelian}
\end{figure}

The diagrammatics are very simple in the abelian CS case. 
The factor of $1/\kappa$ suppresses CS propagators, so the large level limit $\kappa \rightarrow \infty$ 
naively eliminates $\AdS_3$ correlators involving CS lines altogether. However, choosing the normalization for the dual $\CFT_2$ current according to
\begin{align}
j_+ = \kappa \: \underset{\rho \rightarrow 0}{\lim}\: A_+(z^+, z^-, \rho)\:, 
\end{align}
we effectively multiply CS endpoints in $\partial \AdS_3$ Witten diagrams by $\kappa$, canceling the $1/\kappa$ of bulk-boundary propagators, so these survive the limit. Only bulk-bulk CS lines are suppressed.  
The surviving diagrams have the form shown in Fig.~\ref{fig:LargeKappaDiagrams-Abelian}. 

We see that correlators with the $\CFT_3$ are ${\cal O}(\kappa^0)$ (Fig.~\ref{fig:SurvivingDiagram}).  However, the pure CS diagram shown in Fig.~\ref{fig:SpecialDiagram} corresponding to the correlator $\left< j_+ \:j_+ \right>$ is special. While the propagator scales as $1/\kappa$, there are two factors of $\kappa$ for the two end points, making this ${\cal O}(\kappa)$, dominating all other correlators as
 $\kappa \rightarrow \infty$. But, if we restrict our attention to correlations with $\CFT_3$ ``matter'' ($\AdS_4$ particles), then obviously this purely CS correlator drops out and we have a finite limit as $\kappa \rightarrow \infty$.   This explains a puzzle regarding the CS level first seen in $\Mink_4$ AS. For finite but large $\kappa$, $\kappa$ appears in the central extension of the KM algebra as the KM face of the $\left< j_+ \:j_+ \right>$ correlator. 
 But if we are only tracking correlations that involve 4D particles (CFT$_3$), then we are blind to the purely CS correlator 
$\left< j_+ \:j_+ \right>$ and may mistakenly conclude that we are in the limit of vanishing KM level, when in fact we are in the limit of infinite KM level!

\subsection{Non-abelian CS and GR$_3$}

\begin{figure}[h!]
\centering
\begin{subfigure}[b]{.48\textwidth}
  \centering
  \includegraphics[width=.9\linewidth]{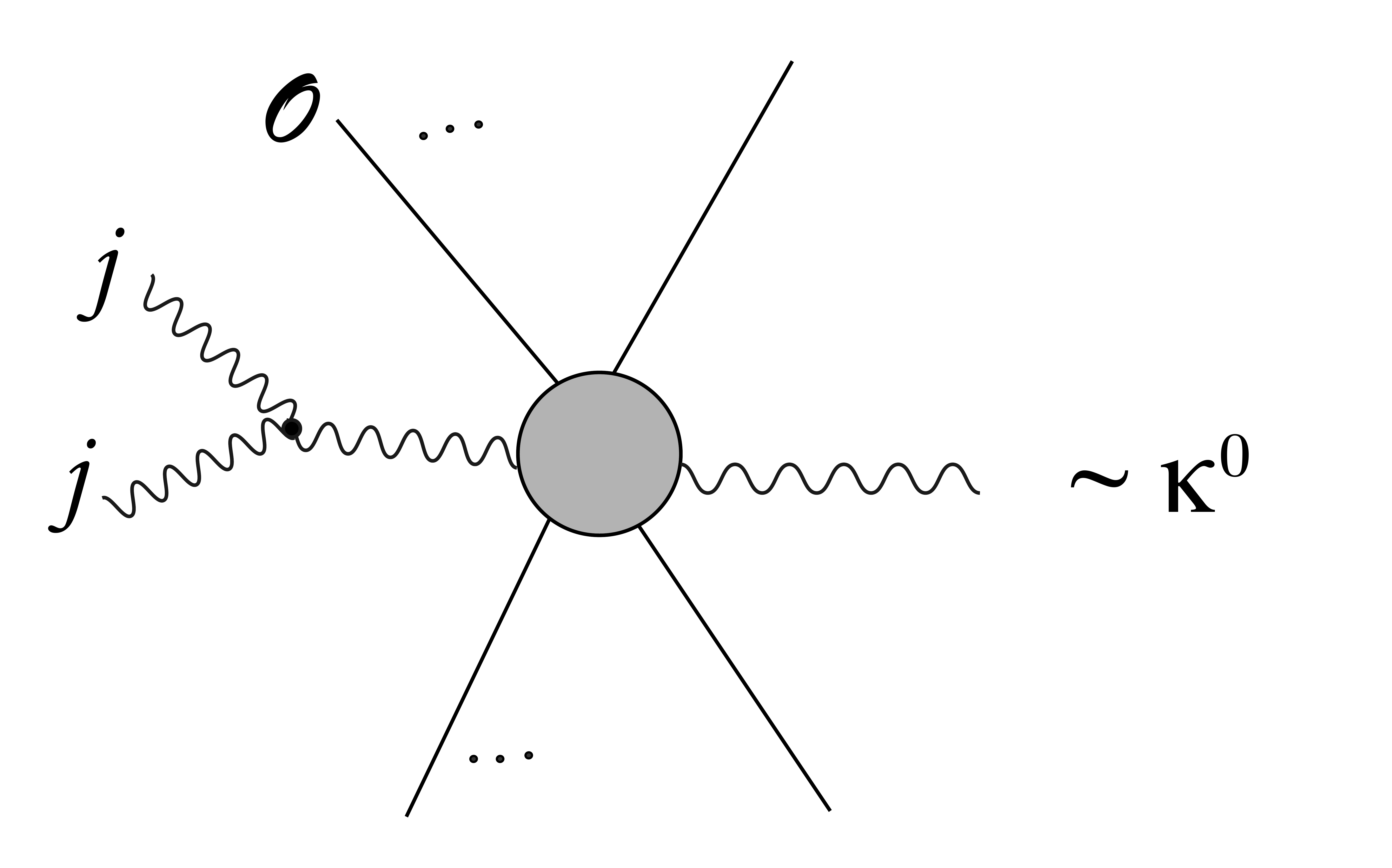}
  \caption{Leading diagrams with external $\CFT_3$ lines.}
  \label{fig:SurvivingDiagram-NonAbelian}
\end{subfigure}\hfill%
\begin{subfigure}[b]{.48\textwidth}
  \centering
  \includegraphics[width=.8\linewidth]{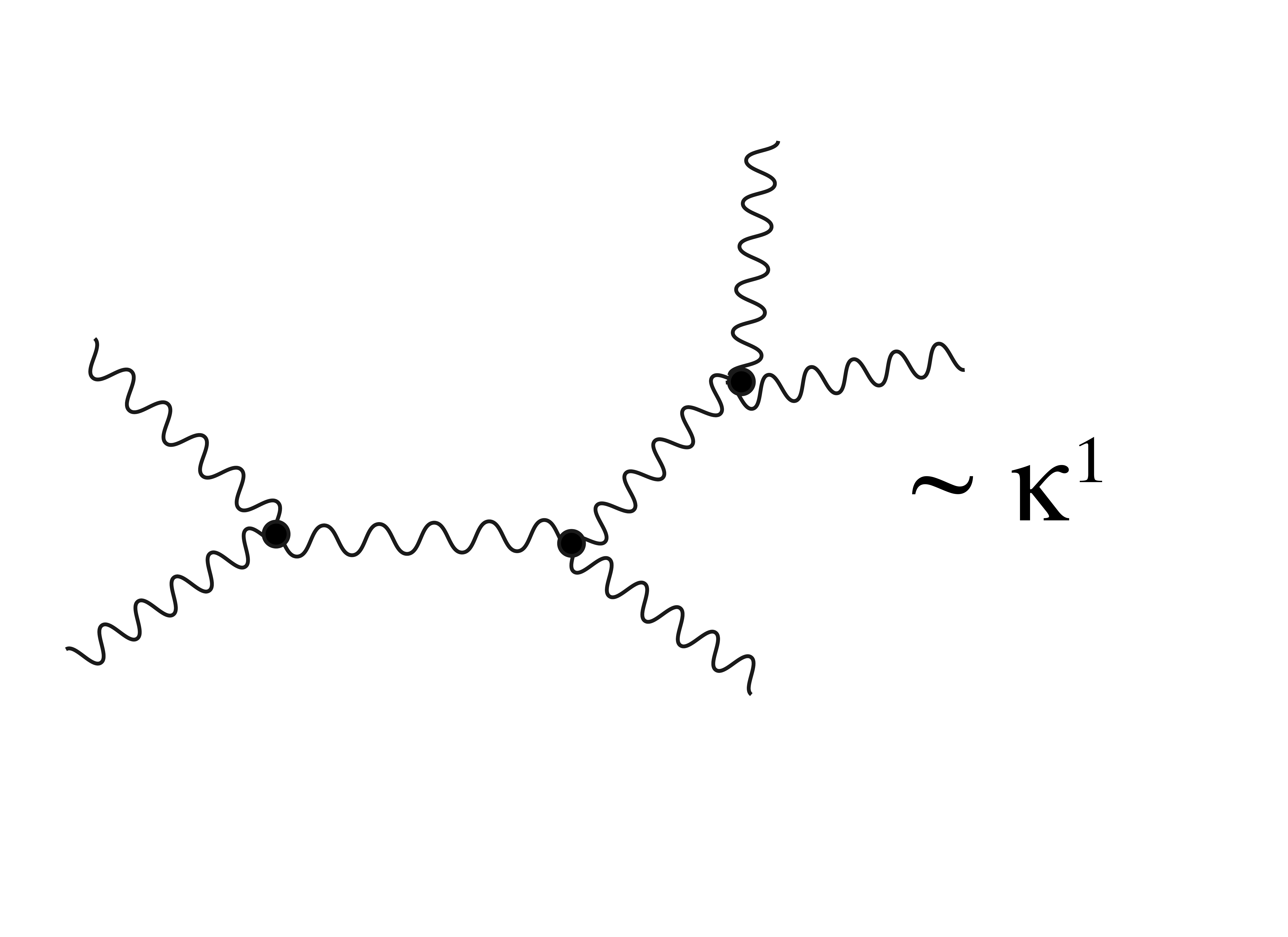}
  \caption{Leading diagrams without external $\CFT_3$ lines.}
  \label{fig:PureCS-NonAbelian}
\end{subfigure}
\caption{Leading non-abelian CS diagrams for large $\kappa$.}
\label{fig:LargeKappaDiagrams-NonAbelian}
\end{figure}

For the case of non-abelian $\text{CS}$ and $\text{GR}_3$, $\kappa \rightarrow \infty$ correlators with $\CFT_3$ hadrons have only tree like CS branches dressing KK $\partial \AdS_3$ Witten diagrams, such as in Fig.~\ref{fig:SurvivingDiagram-NonAbelian}. 
This is very similar to the CS/soft dressing of $\Mink_4$  hard S-matrix elements~\cite{Cheung:2016iub}. While these diagrams are ${\cal O}(\kappa^0)$ for large $\kappa$, again there are ${\cal O}(\kappa)$ correlators given by the pure CS tree diagrams, such as in Fig.~\ref{fig:PureCS-NonAbelian}. And again, focusing on correlations with the CFT$_3$ matter eliminates these, and gives a finite limit as $\kappa \rightarrow \infty$. 

The fact that the CS/GR$_3$ branches attach {\it externally} to CFT$_3$ subdiagrams (blobs), rather than connecting different CFT$_3$ subdiagrams as in Fig.~\ref{fig:TypicalDiagram}, means that the surviving diagrams are effectively purely CFT$_3$ correlators, with the branches just smearing the correlator point for CFT$_3$ currents/stress-tensor where they attach. 
It is these smeared correlators that manifest the CFT$_2$  and AS structure (in large-level limit). That is, in this limit the CS/GR$_3$ are just probes of the dynamical CFT$_3$, with no backreaction on it.   
We discuss the structure and significance of the non-abelian branches as smearing functions in the next section, from the 4D viewpoint.
\section{Non-Standard $\partial \AdS_4/2$ Correlators as CFT$_2$ Correlators}
\label{sec:AdS4-AdS3-Corr}
\begin{figure}[h!]
\centering
\begin{subfigure}[b]{.48\textwidth}
  \centering
  \includegraphics[width=.9\linewidth]{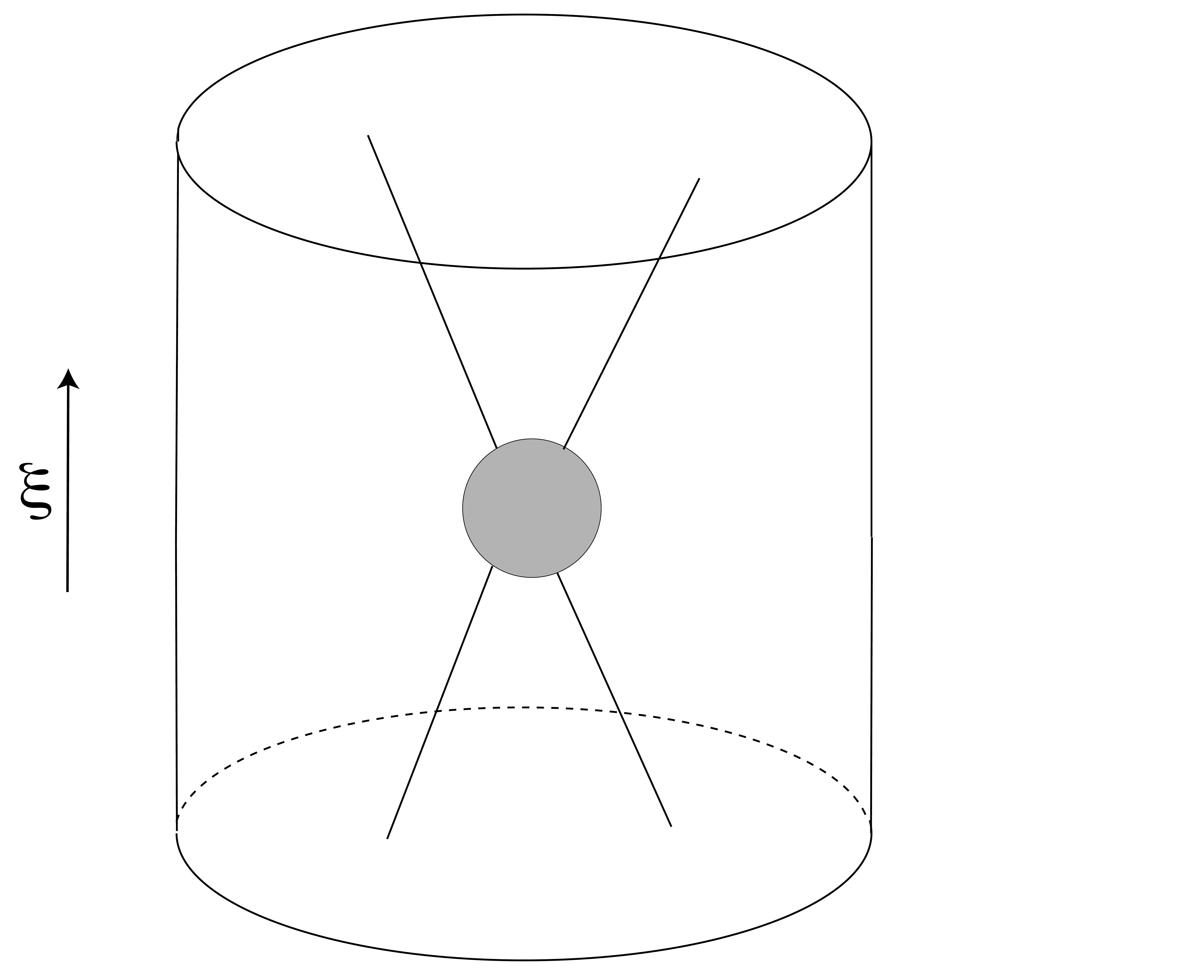}
  \caption{Standard $\partial\AdS_4$ correlators $\neq \partial \AdS_3$ correlators. Here the external lines correspond to superpositions of 3D off-shell ``hadrons''.}
  \label{fig:WittenDiagram-LinesHittingTop}
\end{subfigure}\hfill%
\begin{subfigure}[b]{.48\textwidth}
  \centering
  \includegraphics[width=.9\linewidth]{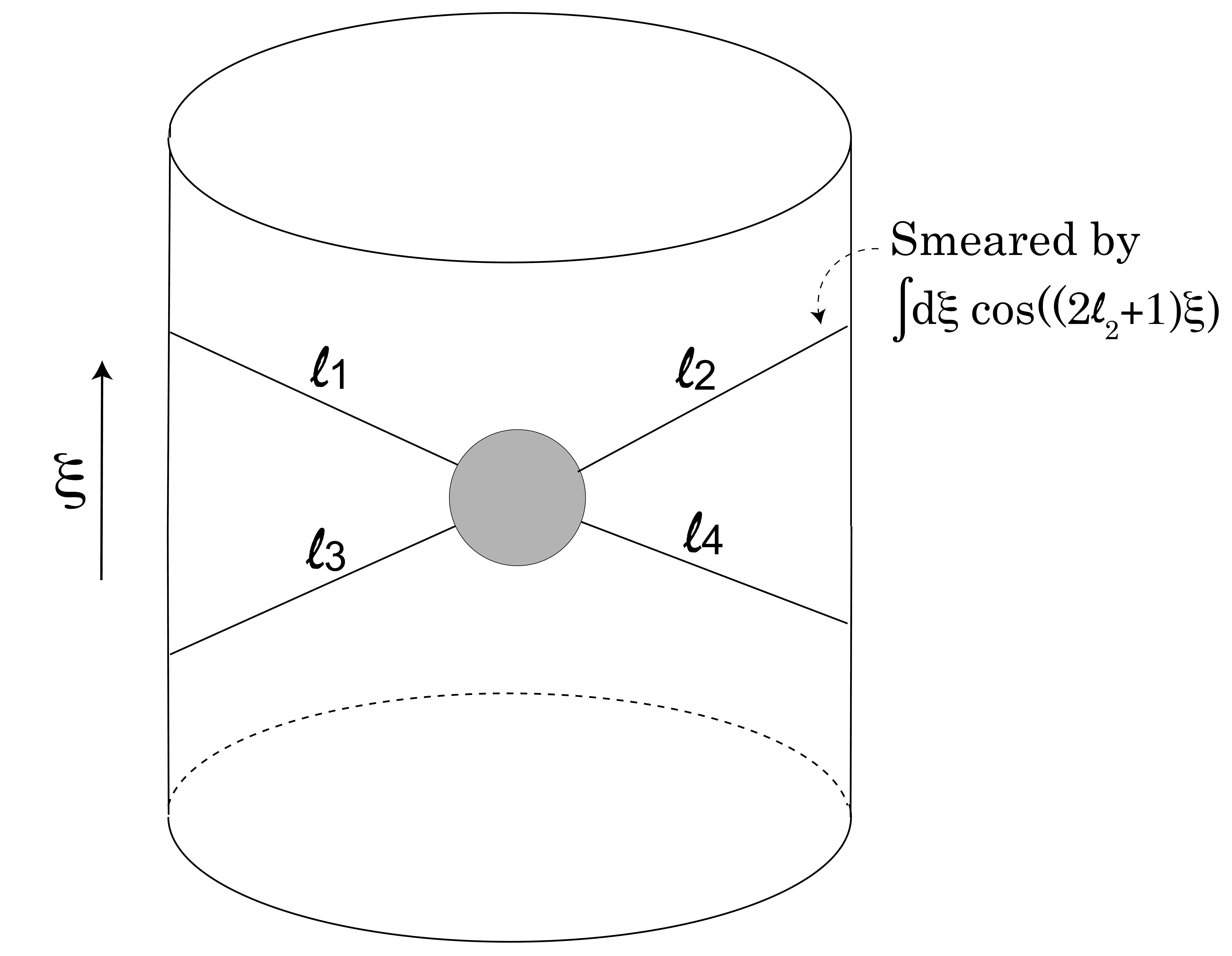}
  \caption{$\partial \AdS_3$ correlator as a non-standard $\partial \AdS_4$ correlator. External lines correspond to 3D on-shell ``hadrons'' (KK modes) of 3D mass $\propto \ell$.}
  \label{fig:WittenDiagram-LinesHittingSide}
\end{subfigure}
\caption{Different types of Witten diagrams in $\AdS_4$ in ``product-space'' coordinates.} 
\label{fig:WittenDiagram-LinesHittingTopAndSide}
\end{figure}

A standard  $\partial \AdS_4$ correlator is  a gauge invariant correlator of local composite operators made of 
 $\CFT_3$ ``quarks'', but from the viewpoint of $\AdS_3$ ``hadron'' mass eigenstates, they are off-shell correlators.
 Instead we are considering  $\partial \AdS_3$ correlators of the ``hadron'' mass eigenstates. In 4D  ``product-space'' coordinates (Eq.~\eqref{eq:AdS4-xiCoordinates}, see Fig.~\ref{fig:AdS4-TinCan}) the distinction is shown in Fig.~\ref{fig:WittenDiagram-LinesHittingTopAndSide}.
These illustrate two alternative means of probing the bulk physics. In standard $\partial \AdS_4$ correlators we are putting sources and detectors on the ceiling and floor of $\AdS_4$ (generic points on the $\partial\AdS_4$ in standard global coordinates) while having signals reflect off the walls with Dirichlet boundary conditions. In the KK-reduced  $\partial \AdS_3$ correlators we have sources and detectors on the walls (only on $\partial\AdS_3 \equiv \text{``equator'' of } \partial\AdS_4$) with signals reflecting off the ceiling and floor with Dirichlet boundary conditions. (In the case of $\AdS_4$/2 we simply put the floor at $\xi = 0$, mid-level in the $\AdS_4$ ``product-space'', with Neumann boundary conditions as discussed earlier.) Either way, no probability or energy is lost through the regions without sources because of the reflecting boundaries.
We stress again that the reason we must insist on the non-standard form of $\partial \AdS_4$ 
correlators is because when CS/GR$_3$ ``emissions'' are added, it is these that become CS/GR$_3$ gauge/diffeomorphism invariant ``on-shell'' $\partial \AdS_3$ correlators. This is in contrast to the non-gauge/diffeomorphism invariance of standard $\partial \AdS_4$ correlators, which are ``off-shell'' from the 
$\AdS_3$ viewpoint. The situation is entirely analogous to the gauge/diffeomorphism invariance of the Minkowski on-shell S-matrix in contrast to the non-invariance of off-shell Minkowski correlators in quantum field theory. 

\subsection{Abelian gauge theory}
\label{subsec:3DGaugeAs4DGauge}

\begin{figure}[h!]
\centering
\includegraphics[width=0.5\textwidth]{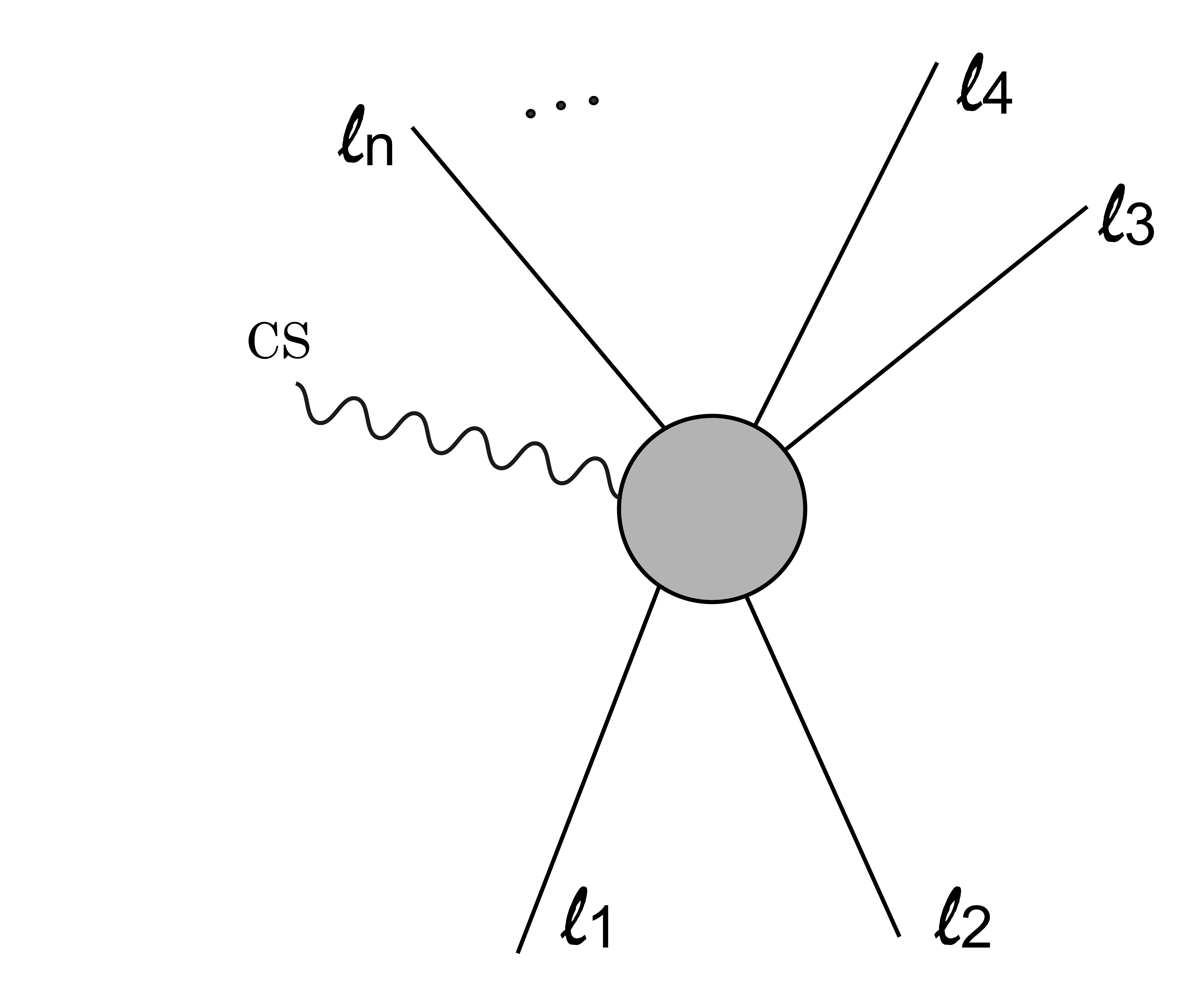}
\caption{3D Witten diagram with external abelian CS line.}
\label{fig:WittenDiagram-OneCSLine}
\end{figure}

\begin{figure}[h!]
\centering
    \includegraphics[width=.5\linewidth]{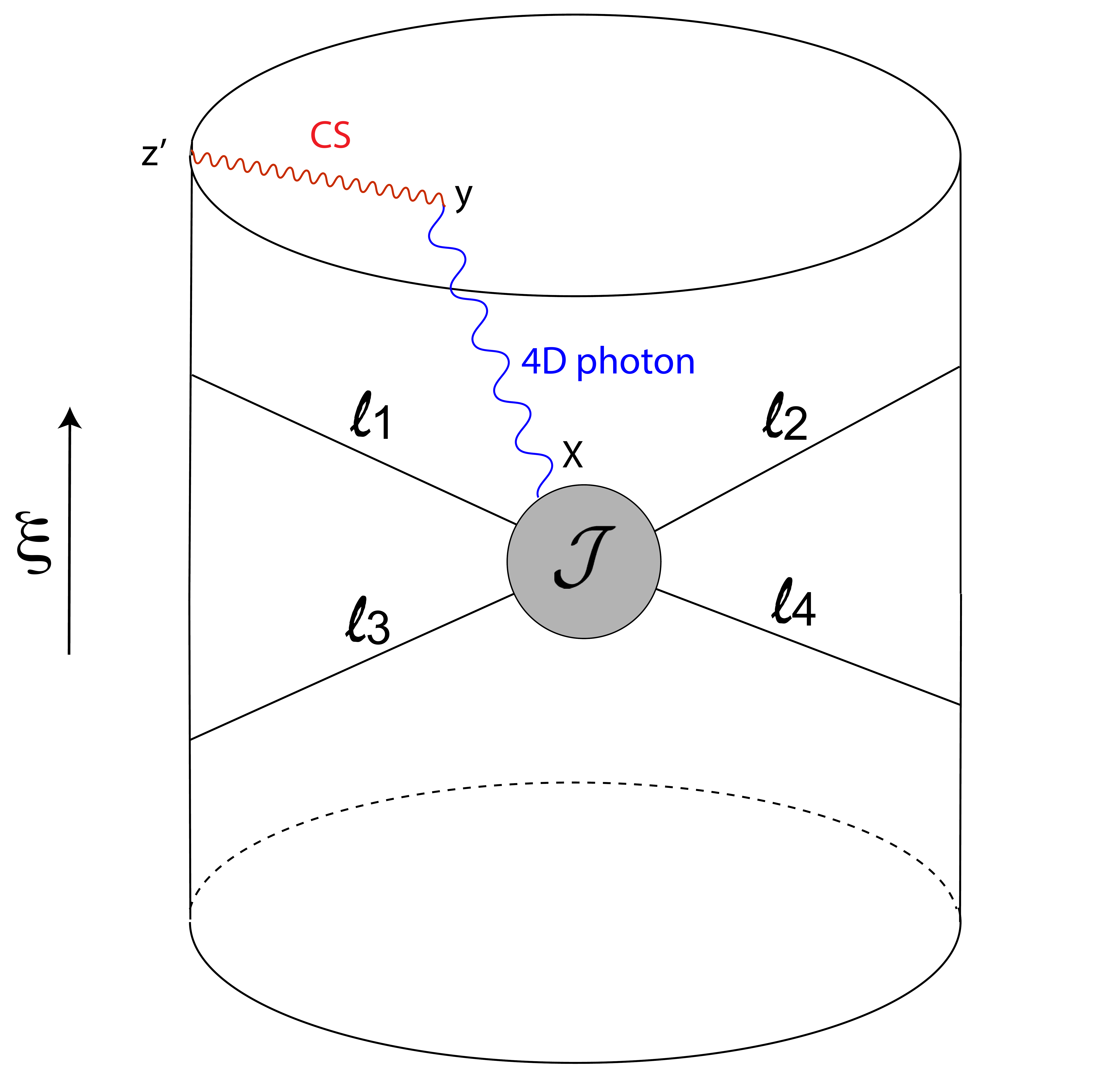}
    \caption{4D Witten diagram in $\AdS_4$, with the external lines ending on $\partial \AdS_3$, including an external 3D CS line (red) ``smearing'' a 4D photon $\partial\AdS_4$ correlator point.}
    \label{fig:WittenDiagram-OneCSLine-2WaysToInterpret}
\end{figure}

For simplicity let us begin by considering $U(1)$ CS coupled to a $U(1)$ symmetry current of CFT$_3$, in turn dual to an AdS$_4$ $U(1)$ gauge field. We focus on 
a 2D chiral current correlator of  $\CFT_2$ with other 2D operators  in the large-$\kappa$ limit. The 2D current of course contains the charges of a $U(1)$ KM algebra by Laurent expansion. 
There are two equivalent ways of reading  such $\CFT_2$ correlators in the large-$\kappa$ limit:
 (i) at face value, as a 3D ``hadronic'' correlator involving CS  ``emission'' (see Fig.~\ref{fig:WittenDiagram-OneCSLine}),  or (ii)  as a purely $\CFT_3$ correlator involving a $\CFT_3$ conserved current at a point $y$ in the $\AdS_3$ {\it bulk} (see Fig.~\ref{fig:WittenDiagram-OneCSLine-2WaysToInterpret}), where this bulk point is ``smeared'' by a function of $y$ given by the $\AdS_3$ CS bulk-boundary propagator:
\begin{equation}
\langle 0 | T \{ j_{+}(z') ... \} |0 \rangle_{\CFT_2} = \int d^3y \sqrt{g_{\AdS_3}} K^{\CS}_{+ \mu}(z', y) 
 \langle 0 | T \{ J_{\CFT_3}^{\mu}(y) ... \} |0 \rangle_{\CFT_3}.
 \end{equation}
By standard AdS$_4$/CFT$_3$ diagrammatics, this lifts to 4D:
\begin{equation}
\langle 0 | T \{ j_{+}(z') ... \} |0 \rangle_{\CFT_2} = \int d^3y \sqrt{g_{\AdS_3}} K^{\CS}_{+ \mu}(z', y) 
\int d^4 X \sqrt{-G_{\AdS_4}} {\cal K}^{\mu}_N (y,X) {\cal J}^{N}(X),
\end{equation}
where ${\cal K}$ is an AdS$_4$ bulk-boundary propagator  corresponding to the 4D photon line in Fig.~\ref{fig:WittenDiagram-OneCSLine-2WaysToInterpret}, while ${\cal J}(X)$ is the bulk 4D current to which it couples, set up by the 4D matter.

We can write this compactly as 
\begin{equation}
\langle 0 | T \{ j_{+}(z') ... \} |0 \rangle_{\CFT_2} = 
\int d^4 X \sqrt{-G_{\AdS_4}} {\cal A}_{N} (X) {\cal J}^{N}(X),
\label{eq:EffectiveAField}
\end{equation}
where 
\begin{equation}
{\cal A}_{N}(X) \equiv \int d^3y \sqrt{g_{\AdS_3}} K^{\CS}_{+ \mu}(z', y) 
 {\cal K}^{\mu}_N(y,X).
 \end{equation}
By the defining properties of  ${\cal K}$ in $\xi$-axial gauge, ${\cal A}_{\mu}(X)$ is that solution to the sourceless 4D Maxwell equations with boundary limit,
\begin{equation}
{\cal A}_{\mu}(y, \xi) 
\underset{\xi \rightarrow \pi/2}{\longrightarrow}
K^{\CS}_{+ \mu}(z', y).
\end{equation}
That is, we deviate from the default Dirichlet boundary condition ${\cal A}_{\mu}=0$ at $\xi = \pi/2$, corresponding to the unperturbed CFT$_3$,  because $K^{\CS}$ acts as a perturbing source for the CFT$_3$ current.

It is straightforward to identify this ${\cal A}$.
Since $K^{\CS}_{+\mu}(z', y)$ is a solution to the free CS equation of motion as a function of $y$, it must be purely a (large) 3D gauge transformation, $K^{\CS}_{+ \mu}(z', y) = \partial_{\mu} \lambda(y)$, specified by its non-trivial boundary limit (at $z'$). This then clearly lifts to the simple 4D solution, 
\begin{equation}
{\cal A}_N(y, \xi) = \partial_N \lambda(y),\qquad {\cal A}_{\mu} = \partial_{\mu} \lambda(y) =  K^{\CS}_{+\mu}(z', y),\qquad {\cal A}_{\xi} = 0.
\label{eq:4DLift}
\end{equation} 
The 3D large gauge transformation of CS is thereby lifted to a large 
4D gauge transformation, such pure gauge configurations being at the root of traditional 4D AS analyses.  Here, substituting Eq.~\eqref{eq:4DLift} into Eq.~\eqref{eq:EffectiveAField} we see that 
\begin{equation}
\langle 0 | T \{ j_{+}(z') ... \} |0 \rangle_{\CFT_2} = \int d^3y \sqrt{g_{\AdS_3}} K^{\CS}_{+ \mu}(z', y) 
{\cal J}_{\eff}^{\mu}(y),
\end{equation}
where 
\begin{equation}
\sqrt{g_{\AdS_3}}  {\cal J}_{\eff}^{\mu}(y) \equiv \int d \xi  \sqrt{-G_{\AdS_4}}     {\cal J}^{\mu}(y, \xi).
\end{equation}
In this way, we see that we can compute $\langle j_+(z') \rangle$ via a CS gauge field coupled either to the holographic CFT$_3$ current $\langle J_{\CFT_3}(y) \rangle$ or the effective ``soft" current made from the 4D bulk, 
 ${\cal J}_{\eff}(y)$.

\subsection{Non-abelian gauge theory and gravity}

\begin{figure}[h!]
\centering
\includegraphics[width=.6\linewidth]{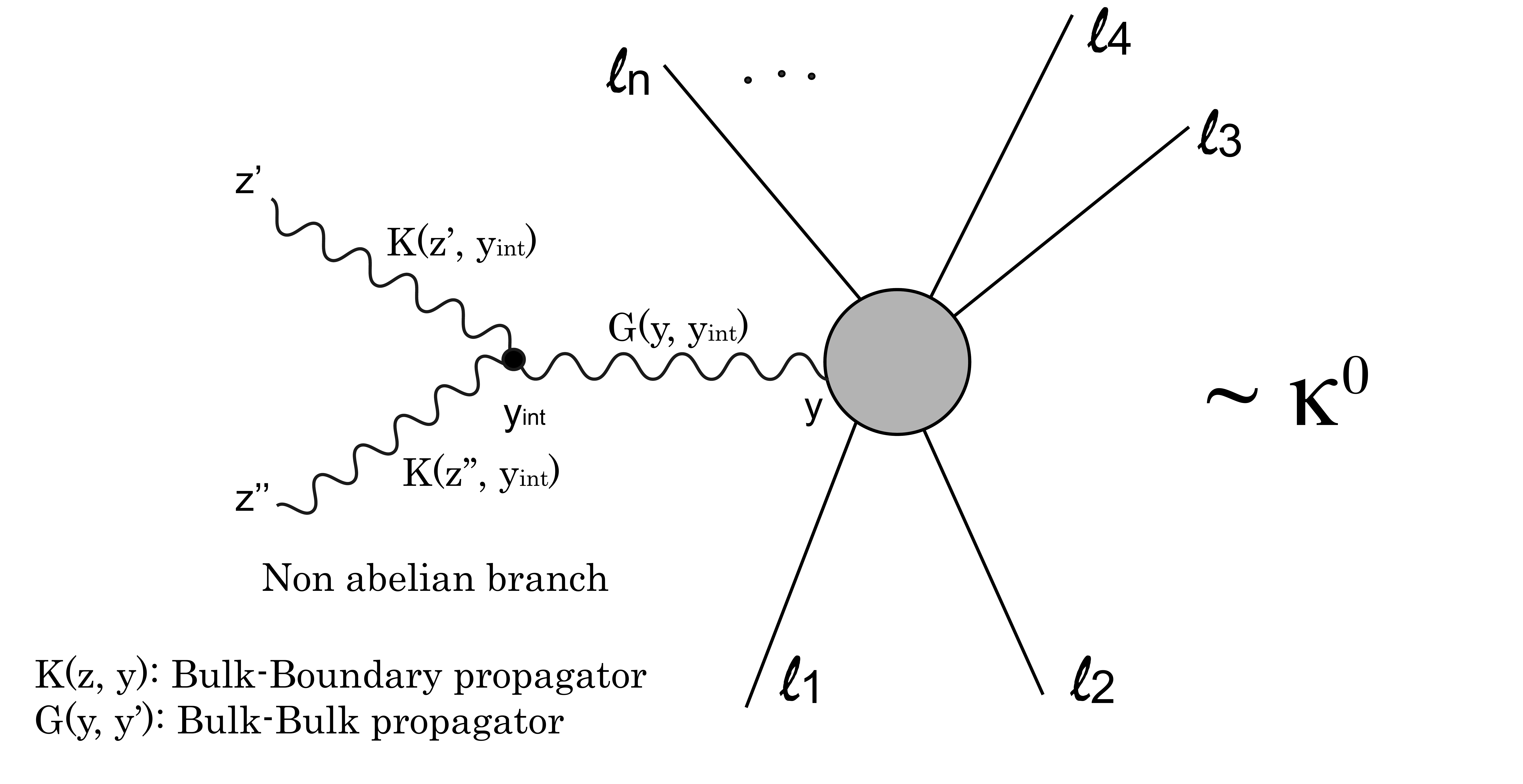}
\caption{Surviving diagram as $\kappa \rightarrow \infty$ with non-abelian CS branch, smearing $\CFT_3$ KK correlator.}
\label{fig:AllNonAbelian}
\end{figure}

Note that in non-abelian gauge theory and gravity, there will be non-abelian CS or $\GR_3$ external branches in correlator diagrams, such as Fig.~\ref{fig:AllNonAbelian}. Again, such correlators can be viewed as purely $\CFT_3$ correlators, but with CFT$_3$ currents/stress-tensor in the $\AdS_3$ bulk, at points $y$ smeared by the non-abelian branch. These branches as functions of $y$ are a non-abelian generalization of abelian CS bulk-boundary propagators, in that they just describe (large) gauge-transformations/diffeomorphisms, because they add up to solutions to the sourceless CS/GR$_3$ equations of motion (with non-trivial boundary limits).
The non-abelian interactions in the branches are just a diagrammatic representation of finding such large gauge-transformations/diffeomorphisms, which is a non-linear problem for non-abelian gauge/diffeomorphism symmetry.
As for the abelian case, these are straightforwardly lifted into 4D large gauge-transformations/diffeomorphisms (as was done in Mink$_4$~~\cite{Cheung:2016iub}).
Thus, once again we see that large gauge-transformations/diffeomorphisms are central to isolating the AS, by suitably smearing $\partial \AdS_4/\CFT_3$ correlators into the canonical form of CFT$_2$ correlators. 

\subsection{Compatibility with 4D quantum loops and masses}

Note that while there are only tree-like CS and $\GR_3$ branches dressing $\CFT_3$/$\AdS_4$ diagrams surviving in the large $\kappa$ limit, the $\CFT_3$ hadron ($\AdS_4$) diagrams can be at full loop level, controlled by a separate parameter such as $1/N_{\CFT_3}$. In this sense, the AS we derive are an all-loop feature, in fact a non-perturbative feature, of $\AdS_4$ QG. Furthermore, while it is technically easier to explicitly consider massless 4D fields, there is absolutely no obstruction to massive 4D fields, dual to high-dimension $\CFT_3$ operators.

\section{Evading the No-Go for Infinite-dimensional AS in $\AdS_4/2$}

\label{sec:Evading-NoGo}

We have derived $\CFT_2$  correlators for 2D currents/stress-tensor in the large level limit from purely  
$\CFT_3$ ($\AdS_4$/2) correlators  of ``hadronic'' (KK) modes and  $\CFT_3$ currents/stress-tensor, smeared by 
large gauge-transformations/diffeomorphisms. This gives rise to infinite dimensional AS of 
 $\VirSquared$ and $\text{KM}$ type. The Virasoro symmetries are analogous to the super-rotations of $\Mink_4$.  
Here, we show from the 4D viewpoint how we have evaded the no-go argument sketched in the introduction for  such infinite-dimensional symmetries of $\AdS_4$, which would equally apply to $\AdS_4/2$. 

 To understand this, note that in ``product-space'' coordinates (Eq.~\eqref{eq:AdS4-xiCoordinates}), there are two distinct $\partial \AdS_4$ regions, the ceiling/floor at $\xi = \pm \pi/2$ and the round wall at $\rho = 0$ (refer to Fig.~\ref{fig:AdS4-TinCan}). 
 These two boundary regions have different conformal structure, 
 \begin{align}
ds^2_{\partial \AdS_4} &\weq
\left\{
  \begin{array}{@{}ll@{}}
    ds^2_{\AdS_3}, & \xi \rightarrow \pm \pi/2 \\
    & \\
    ds^2_{\partial \AdS_3}, & \rho \rightarrow 0. \\
  \end{array}
\right.
\end{align}
In standard global coordinates standard $\partial \AdS_4$ correlators only have sources on the ceiling/floor, and in this boundary region the geometry is fully three-dimensional, with only finite-dimensional conformal isometries as candidate AS.  This is the no-go argument in ``product-space'' coordinates. 
 However, we see that when we put sources only on the wall boundary region, as we have been led to do by the scaffolding of $\GR_3$ and CS on $\AdS_3$, the bulk geometry degenerates as we approach this boundary region to the 2D geometry of $\partial \AdS_3 \equiv S^1 \times \mathbb{R}$, which has infinite-dimensional conformal isometries, corresponding to $\text{Vir}_+ \times {\text{Vir}}_-$ AS.

Thus far $SO(2,2)$ has played the analogous role of Lorentz transformations $SO(3,1)$ in $\Mink_4$, being extended to $\text{Vir}_+ \times \text{Vir}_-$ AS  of $\AdS_4$ analogously to the super-rotations 
$\text{Vir} \times \overline{\text{Vir}}$ of Mink$_4$. The analog of  $\Mink_4$ translation generators are the extra four generators of the $\AdS_4$ isometries 
$SO(3,2)$  which lie outside  $SO(2,2)$, just as Mink$_4$ translations are the Poincare generators outside $SO(3,1)$. We would like to identify these extra generators and the full set of AS of $\AdS_4$ that follows from them, in analogy to XBMS$_4$ incorporating translations to go beyond just the super-rotations in Mink$_4$. 
The problem is that the strategy we used to identify $\CFT_2$ structure forced us to consider $\AdS_4$/2  (rather than $\AdS_4$) and asymptotically $\AdS_3$ GR$_3$, both of which respect only the $SO(2,2)$ subgroup of the global $SO(3,2)$. How can we recover some analog of  ``super-translations'', and more generally the complete AS analog of $\XBMS_4$?

\section{Maximal Spacetime AS from 3D Conformal Gravity}

\label{sec:CGR3}

The infinite-dimensional extension of $SO(2,2)$ isometry arose in our approach by gauging the $\CFT_3$ by $SO(2,2)$ CS =$~\GR_3$ on AdS$_3$. This suggests that we may get the larger infinite-dimensional extension of $SO(3,2)$ by gauging the $\CFT_3$ by $SO(3,2)$ CS instead. Remarkably, this is simply equivalent to 3D conformal gravity ($\CGR_3$)~\cite{Horne:1988jf}.

\subsection{A ``super-translation''-like KM AS for $\AdS_4/2$}

 $\CGR_3$ is compatible with asymptotically $\AdS_3$ spacetime, even though $\AdS_3$ does not have full $SO(3,2)$ conformal isometry. $\CGR_3$ is not only diffeomorphism invariant, but also Weyl invariant. The Weyl invariance shares much in common with an internal $U(1)$ gauge invariance (not coincidentally given Weyl's original gauging of scale symmetry in the history of gauge theory and its similarity to QED's gauging of rephasing invariance). 
Therefore it is not surprising that the AS of $\CGR_3$ (+ $\CFT_3$) matter on $\AdS_3$ are of the form of $\VirSquared$ along with an abelian $\text{KM}$, the latter associated with Weyl symmetry~\cite{Afshar:2011qw}:
\begin{align}
\left[ L_m^+, L_n^+\right] &= (m-n) L^+_{m+n} - \left(\kappa_{\text{grav}} - \frac{1}{12}\right)\left(m^3 - m\right)\:\delta_{m+n,0}\:
\nonumber \\
\left[ L_m^-, L_n^-\right] &= (m-n) L^-_{m+n} + \kappa_{\text{grav}} \left(m^3 - m\right)\:\delta_{m+n,0}\:
\nonumber \\
\left[ L_m^+, J_n^+\right] &= -n J^+_{m+n}\:
\nonumber \\
\left[ J_m^+, J_n^+\right] &= 2\kappa_{\text{grav}} \: m\:\delta_{m+n, 0}\:,
\label{eq:CGR3-algebra}
\end{align}
where $\kappa_{\text{grav}}$ is the level of the $\CGR_3$ theory in CS form, in the same manner as  for GR$_3$. Note, it is critical that the ``quark'' sector is compatible with being gauged by $\CGR_3$, precisely because it is 3D conformally invariant, so that it can be made Weyl-invariant once coupled to gravity. We can interpret the KM resulting from the Weyl invariance as an $\AdS_4$/2 analog of the $\Mink_4$ super-translation KM.

\subsection{Non-unitary nature of $\CGR_3$}

For large $\kappa$ we see that the two Virasoro sub-algebras in
Eq.~\eqref{eq:CGR3-algebra} have opposite sign central charges~\cite{Afshar:2011qw}, $c_- \approx - c_+$, incompatible with unitarity~\cite{Townsend:2013ela}!  This may be surprising because it only pertains to the $\VirSquared $ subalgebra and might be thought to be the same as in $\GR_3$. But crucially, $\GR_3$ is {\it not} a truncation of $\CGR_3$. They employ different quadratic invariants of the generators to define the trace in their CS formulations. Note that for $SO(2,2)$ there are two distinct quadratic invariants, 
\begin{equation}
\epsilon^{IJKL}J_{IJ} J_{KL}\:\:\:\: \text{and} \quad J_{IJ}J^{IJ}\:,\qquad  I,J,K,L = 0, \dots, 3.
\end{equation}
The standard $\GR_3$ formulation uses the first of these and it corresponds to $c_+ = c_-$, so they may both be positive. But the second alternative instead has $c_+ = - c_-$, at odds with that positivity.
For the $SO(3,2)$ CS formulation of $\CGR_3$, there is clearly only a single option,
\begin{equation}
J_{MN}J^{MN}\:, \qquad M,N = 0, \dots, 4,
\end{equation}
and the truncation to $SO(2,2)$ is then the non-positive choice for central charge. 
Nevertheless since we take $\kappa, c \rightarrow \pm \infty$ in our analysis of $\CFT_3$/$\AdS_4$ AS, this does not obstruct the unitarity of the target theory. It does however seem strangely at odds with our development so far, which has made physical sense for finite $\kappa, c$. Possibly, we must restrict to a single CS sector  (4D helicity), say 
``$+$", with $c_+ > 0$~\cite{Townsend:2013ela}.

Although $\CGR_3$ has led us to identify a KM ``super-translation''-like extension of $\AdS_4$/2 AS, this extended algebra still does not contain all of  global $SO(3,2)$, presumably because we are still explicitly breaking  $\AdS_4$ isometries by working with $\AdS_4$/2. We rectify this by first switching to the Poincare patch of AdS$_4$ in the next section,  and then later to all of global AdS$_4$. 

\section{$\AdS_4^\Poincare$:  AS  from Holography and Holography from AS}

\label{sec:PoincarePatch}

We have accumulated a number of questions. Is there an AS algebra of $\AdS_4$ that contains the isometry $SO(3,2)$ as a subgroup? While we are taking the large $\kappa$ limit, what does the finite-$\kappa$ set-up look like in the 4D dual prior to the limit? So far the CS and (C)$\GR_3$ are added ``by hand'', even if then removed by $\kappa \rightarrow \infty$. Is there a sense in which such CS fields emerge as soft limits of the $\AdS_4$ (hence $\CFT_3$) fields themselves, as was the case in $\Mink_4$? If so, do we get a finite emergent level, $\kappa < \infty$? These questions are most simply addressed within the Poincare patch of $\AdS_4$, $\AdS_4^{\Poincare}$. 
The $\AdS_4^{\Poincare}$ metric is given by
\begin{align}
ds^2 = \frac{\eta_{\mu\nu}\:dx^\mu\: dx^\nu - dw^2}{w^2}\:, \qquad 0 < w < \infty\:,
\label{eq:ads-poincare-metric}
\end{align}
which manifests a $\Mink_3$ foliation, where $\eta_{\mu \nu}$ is the $\Mink_3$ metric. Although only a portion of $\AdS_4^{\gl}$, it has the full $\AdS_4$ isometry algebra of $SO(3,2)$, unlike $\AdS_4$/2. We also know its holographic dual, namely $\CFT_3$ on $\Mink_3$, where $SO(3,2)$ are the conformal isometries. 
 
Now we can couple this $\CFT_3$ to $\GR_3$. $\GR_3$ on $\Mink_3$ again has a CS formulation with gauge group $ISO(2,1)$, 
the 3D Poincare group. For finite $\kappa_{\text{grav}}$ the 4D dual of $\GR_3$ $+$ $\CFT_3$ (+ UV completion)  is well known, namely it is the (UV completion of the) Randall-Sundrum 2 (RS2) model~\cite{Randall:1999vf}, but in one dimension lower than the originally formulated~\cite{Emparan:1999wa}. That is, the $\AdS_4^{\Poincare}$ boundary is cut off by a ``Planck brane'' whose 3D geometry is dynamical, dual to $\GR_3$, and coupled to the 4D dynamical bulk geometry (dual to  $\CFT_3$).\footnote{The analogous dual in the case of ${\cal M}_3 = \AdS_3$ is less familiar, a 3D Planck brane in AdS$_4^{\text{global}}$/2.  It is important to distinguish this from the Karch-Randall model~\cite{Karch:2000ct}, in this dimensionality a 3D Planck brane in all of AdS$_4^{\text{global}}$. }

Rather than dwelling on finite $\kappa_{\text{grav}}$, we proceed with the strategy for the 
4D theory to inherit the 3D AS of $\GR_3$ in the large $\kappa_{\text{grav}}$ limit. This AS of Mink$_3$ is $\XBMS_3$~\cite{Ashtekar:1996cd}. Here we review its derivation by a ``contraction'' of the $\VirSquared $ AS of $\AdS_3$, essentially getting flat 3D by taking the $R_{\AdS_3} \rightarrow \infty$ limit~\cite{Barnich:2006av,Bagchi:2009pe,Bagchi:2010eg,Bagchi:2012cy,Barnich:2012aw,Duval:2014uva,Duval:2014lpa}.  

\subsection{XBMS$_3$ from $\VirSquared $}

It is clear in what sense the ``vacuum" geometry of AdS$_3$ approaches Mink$_3$ in the limit of large $R_{\AdS_3}$, 
but we must study the GR$_3$ dynamics as well in this limit in order to understand the relationship of the two AS algebras. GR$_3$ on asymptotically AdS$_3$ can formulated in terms of $SO(2,2) \equiv SO(2,1)_+ \times SO(2,1)_-$ Chern-Simons gauge fields made from the dreibein $e$ and spin connection $\omega$ as~\cite{Witten:1988hc}
\begin{equation}
A_{\mu}^{\pm a} \equiv \omega_{\mu}^{a}  \pm e_{\mu}^{a}/R_{\AdS_3}.
\end{equation}
If we plug this into the AdS$_3$ gravity action in CS form $\equiv S_{CS}(A^+)  - S_{CS}(A^-)$,
and keep the leading terms for large $R_{\AdS_3}$, we find straightforwardly that it is the CS form of the gravity action in Mink$_3$ (with gauge group $ISO(2,1)$) written in terms of $e$ and $\omega$. 

Staying in AdS$_3$, the asymptotic expansion of $A^{\pm}$ in terms of $L^{\pm}$ (reviewed in Section~\ref{sec:CS-review}) 
translates into an expansion for $e$ given by $ R_{\AdS_3} \sum_n (L^+_n - L^-_{-n}) e^{in\phi}$, and for $\omega$ given by  $\sum_n (L^+_n +  L^-_{-n}) e^{in\phi}$.  That is, the AS charges for   $e$ and $\omega$ respectively are
\begin{eqnarray}
R_{\AdS_3} l_n &=& R_{\AdS_3} (L^+_n -  L^-_{-n}) \nonumber \\ 
R_{\AdS_3} T_n &=& (L^+_n +   L^-_{-n}),
\end{eqnarray}
where the overall normalization of $R_{\AdS_3} $ on the left-hand side does not affect relative sizes of terms in the charge algebra, but does give a finite limit as $R_{\AdS_3} \rightarrow \infty$. Indeed, expressing 
the $\VirSquared $ algebra in these variables and taking  $R_{\AdS_3} \rightarrow \infty$ yields the
 centrally-extended X$\BMS_3$: 
\begin{align}
\left[ l_m, l_n\right] &= (m-n) l_{m+n} \nonumber \\
\left[ l_m, T_n\right] &= (m-n) T_{m+n} + \frac{2M_{\text{Pl}}}{12}\: m^3 \: \delta _{m+n, 0}\:,\:\:
\left[ T_m, T_n \right] = 0\:.
\end{align}

As in $\AdS_3$, this X$\BMS_3$ AS is symptomatic of  the topological character of $\text{GR}_3$, the non-trivial topology arising from the ``holes'' drilled out by the matter world lines, where $\text{GR}_3$ reacts by introducing conical-type singularities.

~

We will think of $\XBMS_3$ as an $\AdS_4^{\Poincare}$ analog of ``super-rotations'' in $\Mink_4$ since they are the contraction of $\VirSquared $ AS. The global subalgebra of $\XBMS_3$ is the Poincare isometry $ISO(2,1)$. 
But now  $\AdS_4^{\Poincare}$ ($\Mink_3$) has the larger (conformal) isometry algebra of $SO(3,2)$, containing 
$ISO(2,1)$ as a subalgebra. 
Therefore the extra generators of $SO(3,2)$ can be (repeatedly) commuted with $\XBMS_3$ to generate the full AS of $\AdS_4^{\Poincare}$, with global subgroup $SO(3,2)$! This strategy was analogously followed in $\Mink_4$ as one of the ways to (re-)derive super-translations by commuting ordinary translations with super-rotations~\cite{Cheung:2016iub}.

\subsection{$\CGR_3$ on $\Mink_3$}

Above we outlined a strategy for finding the full AS of $\AdS_4^{\Poincare}$ by starting with its subalgebra, $\XBMS_3$, arising from gauging with $\GR_3$. 
It would be more elegant and insightful if the entire AS emerged by the same procedure. This can now be done by replacing $\GR_3$ by $\CGR_3$ on $\Mink_3$, coupled to $\CFT_3$. Since $\Mink_3$ has $SO(3,2)$ conformal isometries, and $\CGR_3$ is 
$SO(3,2)$ CS, and our ``quark'' matter is also conformally invariant $\CFT_3$, $SO(3,2)$ is respected by each component, and therefore the infinite dimensional symmetries that arise from the CS structure must contain all of $SO(3,2)$ as a global subalgebra. We will pursue the explicit form of this AS algebra elsewhere, just observing here that it is implicitly completely characterized by CGR$_3$ on Mink$_3$.

\subsection{Holographic Grammar from AS}

While we have used the holographic grammar of $\AdS_4$/$\CFT_3$ in this paper to clarify the nature and utility of AS, we can run our arguments in a different order. Suppose that one did not know the holographic dual of $\AdS_4$ QG, but was given the full AS structure of $\AdS_4$ and learned to characterize it in terms of $\CGR_3$ fields to capture the associated large gauge transformations. Then by the fact that matter compatible with coupling to 3D gravity must be a 3D local quantum field theory in order to have the requisite local stress tensor to source gravity, we can deduce that the holographic dual of $\AdS_4$ must be such a 3D QFT. The fact that the 3D gravity is specifically conformal gravity implies that the dual 3D QFT must also be conformally invariant, that is $\CFT_3$! It is just such a set of steps that awaits to be performed in the case of finding a holographic grammar behind $\Mink_4$ QG. 

\section{Emergent CS and ``Shadow'' Effects from Boundary/Soft Limits} 
\label{sec:Emergent}

In $\Mink_4$ gauge theory, it was shown that AS and memory effects arise from considering same-helicity gauge boson emissions in the soft limit \cite{He:2014cra,He:2015zea}. Ref.~\citep{Cheung:2016iub} showed that these features were captured by an emergent 3D CS description of the soft fields, ``living" at $\partial \Mink_4$, as well as on Rindler/Milne horizons. Here, we will demonstrate that analogous phenomena emerge within AdS$_4^{\Poincare}$ $U(1)$ gauge theory. (If 
AdS$_4^{\Poincare}$ $\GR_4$ is added, we can think  of these phenomena as emerging from within the dual $\CFT_3$ with  $U(1)$ global symmetry, even though the gravity will play no explicit role in our analysis.) 
While $\AdS_4^{\gl}$ has a  discrete spectrum, AdS$_4^{\Poincare}$  has a continuous spectrum and a natural generalization of ``soft" limit. 
We find emergent CS gauge fields localized on $\partial$AdS$_4^{\Poincare}$ as well as on the Poincare horizon, connected by this soft limit. These CS fields connect to analogs of electromagnetic memory effects in Mink$_4$ \cite{Bieri:2013hqa,Pasterski:2015zua,Susskind:2015hpa,Cheung:2016iub}, which we will refer to as ``shadow" effects, since they relate to the holographically emergent spatial direction rather than time. We will also see a sense in which a finite CS level emerges. While our approach here parallels similar steps in Mink$_4$ \cite{Cheung:2016iub}, the emergent CS structure in $\AdS_4$ is closely related to ``mirror" symmetry in the dual $\CFT_3$ \cite{Intriligator:1996ex,Kapustin:1999ha,Witten:2003ya}. This aspect will be explored elsewhere~\cite{InPrep}. 

\subsection{Set-up}

We consider an  AdS$_4^{\Poincare}$ $U(1)$ Maxwell gauge field ${\cal A}_{N}$,  
coupled to a bulk 4D conserved source current ${\cal J}_N$, which is taken to implicitly describe interacting charged matter. The 4D gauge coupling is $g$. 
Because of the Weyl invariance of the 4D Maxwell action, $\AdS_4^{\Poincare}$ (Eq.~\eqref{eq:ads-poincare-metric})
is effectively just $\Mink_4$/2, 
\begin{align}
ds^2 \weq \eta_{\mu\nu}\:dx^\mu\:dx^\nu - dw^2\:, \qquad w > 0,\:\mu,\nu = \{0,1,2\}.
\end{align}

The natural notion of ``soft''  in $\AdS_4^{\Poincare}$/$\CFT_3$ is $m_3^2 \rightarrow 0$, where $m_3^2$ is 
3D invariant mass-squared in the $x^{\mu}$ directions. This is  
 obviously analogous to the observation of Ref.~\cite{Cheung:2016iub} 
that $\Mink_4$ soft limits correspond to $m_3 \rightarrow 0$ in the  $\text{(EA)dS}_3$ foliation of $\Mink_4$. 

Maxwell radiation  can be decomposed into positive and negative helicity components, 
${\cal A}^{\pm}$. More generally, away from charged matter (away from the support of ${\cal J}$), 
we will decompose the electromagnetic field strength ${\cal F}_{MN}$
 into self-dual and anti-self-dual components, 
\begin{equation}
  {\cal F}^\pm_{MN}(x, w =0) 
  \equiv 
  \frac{1}{2}\left({\cal F}_{MN} \pm i \widetilde{\cal F}_{MN} \right)
  \equiv \partial_{M} {\cal A}^{\pm}_{N} -  
\partial_{N} {\cal A}^{\pm}_{M}.
\end{equation}
 We will focus on the soft limit  of ${\cal A}^+$. 
Let us first imagine that we are in full Mink$_4$ instead of Mink$_4/2$. In momentum space, $(q_{\mu}, q_w)$, $m_3^2 = q_{\mu} q^{\mu}$, so that for 4D on-shell radiation, $m_3^2 = q_w^2$. More precisely, the {\it leading} soft limit would be given by
\begin{equation}
\lim_{q_w \rightarrow 0} q_w \: {\cal A}^+(q_w) = \int_{- \infty}^{\infty} dw \: \partial_w {\cal A}^+(w),
\end{equation}
where the 3D argument is implicit and can be either $q_{\mu}$ or $x^{\mu}$. 
Within Mink$_4/2$, the analogous soft limit is truncated to\footnote{We can think of Mink$_4/2$ as the quotient space of Mink$_4$ under the identification $w \leftrightarrow - w$. If we imagine a  ``polarizer" projecting
onto positive helicity in the physical region, $w > 0$, and its ``mirror image" projecting onto negative helicity for $w < 0$,
then the definition of leading soft limit in the Mink$_4$ covering space reduces to the truncated expression in Mink$_4/2$.}
\begin{equation}
\int_{0}^{\infty} dw \: \partial_w {\cal A}^+(w) = {\cal A}^+(w = \infty) -  {\cal A}^+(w = 0). 
\end{equation}
We will take this as our ``soft limit".

In what follows, we will see that each of the 3D fields in this soft limit, ${\cal A}^+(x, w = \infty)$ and ${\cal A}^+(x, w = 0)$ obeys an interesting CS-type equation.

\subsection{CS on $\partial \AdS_4^{\Poincare}$ and a ``holographic shadow" effect} 

Consider that the source current ${\cal J}$ emits radiation towards $\partial \AdS_4^{\Poincare}$. The positive helicity component at $\partial \AdS_4^{\Poincare}$ satisfies
\begin{equation}
{\cal F}^+_{\mu \nu}(x, w =0) =  \frac{i}{2} \widetilde{\cal F}_{\mu \nu}(x, w =0) \equiv -\frac{i}{4} \epsilon_{\mu \nu \rho} {\cal F}^{w \rho}(x, w =0),
\end{equation}
because the standard $\AdS_4$ Dirichlet boundary condition, ${\cal A}_{\mu}(x, w=0) = 0$, implies
${\cal F}_{\mu \nu}(x, w=0) = 0$. 
In terms of the standard AdS$_4$/CFT$_3$ dictionary for the holographic symmetry current, 
\begin{equation}
J_{\CFT_3}^{\rho}(x) = \frac{1}{g} {\cal F}^{w \rho}(x, w=0),
\end{equation}
we obtain
\begin{equation}
{\cal F}^+_{\mu \nu}(x, w =0) =  - \frac{i}{4} g \: \epsilon_{\mu \nu \rho} J_{\CFT_3}^{\rho}(x).
\end{equation}

We can view this as the equation of motion for an emergent CS gauge field coupled to CFT$_3$ charged matter, 
\begin{equation}
 F^{\CS}_{\mu \nu}(x) =  g\: \epsilon_{\mu \nu \rho} J_{\CFT_3}^{\rho}(x),
\end{equation}
where the CS gauge field is identified with the helicity-cut boundary limit of the 4D gauge field in $w$-axial gauge, 
\begin{equation}
A^{\CS}_{\mu}(x)  \equiv 4i {\cal A}^+_{\mu}(x, w=0).
\end{equation}
(This does not vanish since only 
${\cal A} = {\cal A}^+ + {\cal  A}^-$ obeys the AdS Dirichlet boundary condition.)
It was just such a CS field coupled to CFT$_3$ (but on AdS$_3$ instead of Mink$_3$) which was invoked in earlier sections  to derive AS for AdS$_4$. 

It is useful to cast  the CS equation in integrated form, using Stokes' Theorem,
\begin{equation}
\oint_{\partial \Sigma} dx^{\rho} A^{\CS}_{\rho}(x) = g \int_{\Sigma} d^2 \Sigma^{\mu \nu} \epsilon_{\mu \nu \rho} J_{CFT_3}^{\rho}.
\end{equation}
Here $\Sigma$ is a finite two-dimensional surface in the  $\partial$AdS$_4^{\Poincare}$ $x$-spacetime, with boundary $\partial \Sigma$. For example, for purely spatial $\Sigma$, the right-hand side is the total CFT$_3$ ``quark" charge lying inside $\Sigma$, a holographic ``shadow" of the 4D bulk state.

\subsection{Emergent CS level}

As explained earlier, in CS theory, sensitivity to the CS level $\kappa$ (in correlators with external matter lines) arises from diagrams with internal CS lines. In the present context, we have considered radiation emitted by a source ${\cal J}$. The CS gauge field is the boundary limit of the positive helicity component of this 4D radiation, ${\cal A}^+(w=0)$. To measure the associated CS level we imagine ``detecting" this field with a probe charge localized near or at the boundary, $w=0$. 

The subtlety is that physical charges couple to both positive and negative helicity components. We straightforwardly see that the boundary limit of the negative helicity component ${\cal A}^-(w=0)$ satisfies
\begin{equation}
{\cal F}^-_{\mu \nu}(x, w =0) =  \frac{i}{4} g \: \epsilon_{\mu \nu \rho} J_{\CFT_3}^{\rho}(x).
\end{equation}
That is, while the probe charge couples to the sum of the helicity components in the form, $g {\cal A} = 
g {\cal A}^+ + g {\cal A}^-$, the two helicities couple with {\it opposite} strength to the holographic current, in the form $\pm g J _{\CFT_3}$.  Therefore the CS exchanges mediated by ${\cal A}^+$ and ${\cal A}^-$ have strengths 
$\pm g^2$, yielding a net cancelation.  However, we can formally focus on just the ${\cal A}^+(w=0)$ CS exchange with strength
$+ g^2$, corresponding to CS level, 
\begin{equation}
\kappa \sim \frac{1}{g^2}.
\end{equation}
A similar result was anticipated in Ref.~\cite{Cheung:2016iub} for Mink$_4$.

\subsection{The soft limit, CS on the Poincare horizon, and a bulk ``shadow" effect} 
\label{subsec:soft-limit}

The $\mu$-component of the 4D Maxwell equations reads
\begin{equation}
\partial_w {\cal F}^{w \mu} + \partial_{\nu} {\cal F}^{\nu \mu} = g {\cal J}^{\mu}.
\end{equation}
We again look at an integrated form of these equations, on a two-dimensional surface 
 $\Sigma$ in  $\partial$AdS$_4^{\Poincare}$ $x$-spacetime, and in our ``soft limit" in $w$. 
 That is, we integrate with respect to the three-volume,
$\int_0^{\infty} dw \int d^2 \Sigma^{\rho \sigma} \epsilon_{\rho \sigma \mu}$ ...~, to get
\begin{eqnarray}
\int d^2 \Sigma^{\rho \sigma} \epsilon_{\rho \sigma \mu} 
\Big[ {\cal F}^{w \mu}(w=\infty) -  {\cal F}^{w \mu}(w=0) \Big] 
&~& \nonumber \\ 
 + \int_0^{\infty} dw \oint_{\partial \Sigma} dx^{\rho} \epsilon_{\rho \mu \nu} {\cal F}^{\mu \nu} &=& 
g  \int_0^{\infty} dw \int d^2 \Sigma^{\rho \sigma} \epsilon_{\rho \sigma \mu} {\cal J}^{\mu},
\end{eqnarray}
where we have used Stokes' Theorem to get the second line of the left-hand side. 
For the simple case of purely spatial $\Sigma$ this is nothing but Gauss' Law, the right-hand side being just the total bulk charge lying inside the three-volume, while the left-hand side is the total electric flux through its boundary. 

Taking the source current ${\cal J}$ to be localized at finite $w$ at finite times, and $\Sigma$ to only span finite times, we can drop the field strength at $w= \infty$ on the first line, by causality. The field strength at $w=0$ on the first line is just the holographic current again, so we have 
 \begin{equation} 
 \int_0^{\infty} dw \oint_{\partial \Sigma} dx^{\rho} \epsilon_{\rho \mu \nu} {\cal F}^{\mu \nu}
= g \int d^2 \Sigma^{\rho \sigma} \epsilon_{\rho \sigma \mu} \Big[J_{\CFT_3}^{\mu}  + \int_0^{\infty} dw {\cal J}^{\mu}(w)\Big]. 
\end{equation}
This is closely analogous to the electromagnetic memory effect in Mink$_4$ for purely spatial $\Sigma$, with $w$ now playing the role of time there. The total charge passing through $\Sigma$ {\it regardless of when}
in the memory effect is replaced here by the total charge in $\Sigma$ {\it regardless of where in} $w$.
We will refer to this as a ``bulk shadow" effect. 

As was done for the memory effect in Ref.~\cite{Cheung:2016iub}, we can write the bulk shadow effect in CS form. First note that 
the left-hand side can be re-expressed in terms of the dual field strength $\widetilde{\cal F}$ to give
\begin{equation} 
2 \oint_{\partial \Sigma} dx_{\rho}  \int_0^{\infty} dw \widetilde{\cal F}^{w \rho} 
= g \int d^2 \Sigma^{\rho \sigma} \epsilon_{\rho \sigma \mu} \Big[J_{\CFT_3}^{\mu}  + {\cal J}_{\eff}^{\mu}\Big],
\end{equation}
where we have defined a second 3D  ``shadow" current by taking the soft limit of the bulk 4D current,
\begin{equation}
{\cal J}_{\eff}^{\mu}(x) \equiv \int_0^{\infty} dw  {\cal J}^{\mu}(x, w). 
\end{equation}
We add zero to the bulk shadow effect in the form,
\begin{eqnarray}
0 &=& 2 i \oint_{\partial \Sigma} dx_{\rho}  \int_0^{\infty} dw {\cal F}^{w \rho} \nonumber \\
&=&  2 i \oint_{\partial \Sigma} dx_{\rho} {\cal A}^{\rho}(w=\infty) -  2 i \oint_{\partial \Sigma} dx_{\rho} {\cal A}^{\rho}(w=0),
\end{eqnarray}
where the term at 
$w = \infty$ is by Stokes' Theorem $ = i \int d^2 \Sigma_{\mu \nu} {\cal F}^{\mu \nu}(w=\infty)$, which  vanishes by causality, and the term at $w=0$ vanishes by the standard AdS$_4$ Dirichlet boundary conditions on ${\cal A}$.
 Therefore we can write the bulk shadow effect in the form
\begin{equation} 
- 2 i \oint_{\partial \Sigma} dx_{\rho}  \int_0^{\infty} dw \big[{\cal F}^{w \rho}  + i \widetilde{\cal F}^{w \rho}\big]
= g \int d^2 \Sigma^{\rho \sigma} \epsilon_{\rho \sigma \mu} \big[J_{\CFT_3}^{\mu}  + {\cal J}_{\eff}^{\mu}\big].
\end{equation}

It is straightforward for $\partial \Sigma$ to avoid the support of ${\cal J}$  for all $w$, so that the self-dual component of the field strength on the left-hand side can be expressed in terms of the gauge potential ${\cal A}^+$. By Stokes' Theorem, 
\begin{equation} 
 \oint_{\partial \Sigma} dx^{\rho}  {\cal A}^+_{\rho}(w=\infty) - 
\oint_{\partial \Sigma} dx^{\rho}  {\cal A}^+_{\rho}(w=0)
= \frac{ig}{4} \int d^2 \Sigma^{\rho \sigma} \epsilon_{\rho \sigma \mu} \Big[J_{\CFT_3}^{\mu}  + {\cal J}_{\eff}^{\mu}\Big].
\end{equation}
We see that the term on the left at $w=0$ and the $J_{\CFT_3}$ term on the right are equal by the last subsection, so we isolate a new CS-type relation on the Poincare horizon, 
\begin{equation} 
 \oint_{\partial \Sigma} dx^{\rho}  {\cal A}^+_{\rho}(w=\infty)  
= \frac{ig}{4}  \int d^2 \Sigma^{\rho \sigma} \epsilon_{\rho \sigma \mu}  {\cal J}_{\eff}^{\mu}.
\end{equation}
This is the ($\Sigma$-integrated) CS form of the bulk shadow effect, where the role of CS current is
played by the shadow current, ${\cal J}_{\eff}$. 

In subsection~\ref{subsec:3DGaugeAs4DGauge}, with CFT$_3$ on AdS$_3$, we saw that AS ($\CFT_2$ chiral current $j_+$) could be derived by CS coupled to either $J_{\CFT_3}$ {\it or} 
${\cal J}_{\eff}$.  But this equivalence required going to the boundary of AdS$_3$. For $\Sigma$ in the ``bulk" of 
Mink$_3$, the two CS relations at $w=0$ and $w=\infty$, with CS currents 
$J_{\CFT_3}$ and  
${\cal J}_{\eff}$ respectively, are distinct. 

In the same sense as for the CS gauge field localized at the boundary, the CS gauge field on the Poincare horizon
also has level $\kappa_{\eff} \sim 1/g^2$.

\section{AS of Wheeler-DeWitt Wavefunctionals on $\partial \AdS^{\gl}_4$}

\label{sec:s2-times-R-cutInTime}

The choices of ${\cal M} = \AdS_3$ and $\Mink_3$ have given an approach to AS on portions of $\AdS_4^{\gl}$, but here we return to the full $\AdS_4^{\gl}$. It is natural then to consider 
 CS coupled to $\CFT_3$ on the global boundary $S^2 \times \mathbb{R}$. However, space is then closed and there is no obvious asymptotic region to get AS or 2D chiral currents. 
Yet, it is well known from the CS viewpoint that there are effectively infinite-dimensional symmetries still at play, and these are revealed by cutting at a time slice to reveal a state~\cite{Witten:1988hf}. Technically, this is clear if we consider the wavefunctional, say at time $\tau =0$, to be determined by a 3D functional integral over all earlier times $\tau < 0$ and  all of space, that is effectively ${\cal M}_3 \equiv S^2 \times \mathbb{R}^{-}$, where $\mathbb{R}^{-}$ is the negative-$\tau$ half-line. Once again, this spacetime has a boundary, the $S^2$ space at $\tau = 0$, on which AS appear in standard CS fashion. They act on the states of the theory.

Let us return to the no-go argument for infinite-dimensional AS of AdS$_4$ and the loop-hole pointed out in the introduction. 
CFT$_3$ states are dual to $\AdS_4$  diffeomorphism-invariant Wheeler-DeWitt wavefunctionals.
In particular they describe the state at $\tau = 0$ on the boundary, but on any interpolating spacelike hypersurface in the bulk.  The collection of such hypersurfaces gives the 4D subregion of AdS$_4$ described by the quantum state, as depicted in Fig.~\ref{fig:Wheeler-DeWitt-AdS4}. Its  boundary geometry is effectively two-dimensional, compatible with infinite dimensional AS. 

\begin{figure}[h!]
\centering
\includegraphics[width=0.45\textwidth]{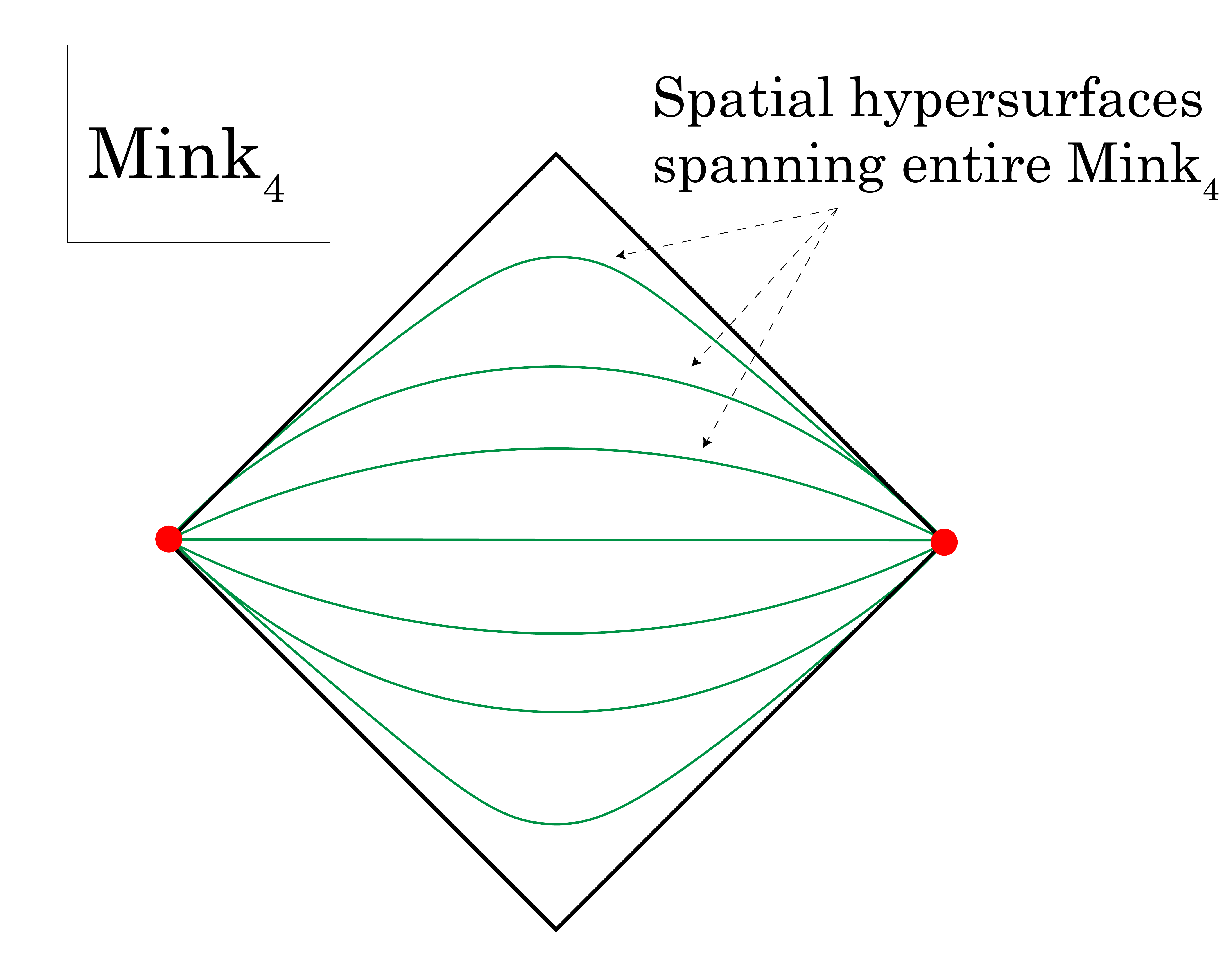}
\caption{In asymptotic $\Mink_4$ spacetime, different time-slices (given by green spatial hypersurfaces) are just related by diffeomorphisms within a single ``timeless'' Wheeler-DeWitt wavefunctional.}
\label{fig:Wheeler-DeWitt-Mink4}
\end{figure}

Such a restriction to a subregion does not occur for  $\Mink_4$. The quantum state 
 at Minkowski time $= 0$ on the boundary describes the 4D region foliated by all interpolating spacelike hypersurfaces, as for $\AdS_4$, but unlike $\AdS_4$ this foliation covers {\it all} of Mink$_4$. See Fig.~\ref{fig:Wheeler-DeWitt-Mink4}.

\subsection{CS gauge theory on $\partial \AdS^{\gl}_4$}

Consider a $U(1)$ CS field for simplicity. The CS field sees the $U(1)$ charged CFT$_3$ state at $\tau = 0$ via an Aharonov-Bohm(AB) phase in Wilson loops.
 One can define associated charges,
\begin{equation}
Q_{\Sigma} \equiv  \oint_{\partial \Sigma} d \ell \cdot A,
\end{equation}
measuring the total ``quark" charge inside subregion $\Sigma$ of the spatial $S^2$ at $\tau = 0$, 
by the integrated form of equations of motion for CS coupled to CFT$_3$. 

These contour-associated AS charges are related to the standard KM charges as follows. We use complex coordinates $z, \bar{z}$ on $S^2$ via sterographic projection. 
Out of the two boundary components of the CS gauge field, $A_z, A_{\bar{z}}$, one component is removed by a CS boundary condition, say $ A_{\bar{z}} = 0$, while the other component is holomorphically conserved, 
$\partial_{\bar{z}} A_z = 0$ (refer to Section~\ref{sec:CS-review}). This holomorphic $A_z(z)$ is then completely determined by the poles at the location of charged $\CFT_3$ ``quarks'', so that 
\begin{equation}
Q_{\Sigma} \equiv  \frac{1}{2 \pi i} \oint_{\partial \Sigma} d z A_z(z) 
\end{equation}
as a complex contour integral. Laurent expanding about $z=0$ say, 
\begin{equation}
A_z(z) = \sum_n \frac{Q_n}{z^{n+1}}, 
\end{equation}
then determines the KM charges. 
%
%
%
%
%
Note that even as the CS is decoupled at $\kappa = \infty$, the $Q_{\Sigma}, Q_n$ remain as non-gauged charges registering the location of charged ``quarks'', and therefore the holographic boundary ``shadows'' of 4D particles. 

~

While this form of  ``hair" for 4D charges is amusing, the key question is whether it is {\it useful}, say in the sense that it makes time-evolution algebraic in terms of the charges, as opposed to having to solve complicated dynamics. 
We have already seen how such simple time-evolution of charges arises for ${\cal M} = \AdS_3$ in the context of $\AdS_4/2$ (see Eq.~\eqref{eq:KM-Charge-Tau-Dependence}). 
To see the analogous form of time-evolution of charges 
 in the $S^2 \times \mathbb{R}$ setting we need the full power of CGR$_3$.

\subsection{Time-evolution from AS algebra via $\CGR_3$ on $\partial \AdS_4^{\gl}$}

\label{sec:s2-times-R}

Given the CS form of 3D gravity, one might think to just repeat the above steps performed for internal CS gauge symmetries. 
But now $S^2 \times \mathbb{R}$ geometry must be a solution to dynamical gravity. 
And yet, for standard GR$_3$ (with or without a cosmological constant) it is not a solution to 3D Einstein equations.
The closest is $\GR_3$ with positive cosmological constant, which has $\dS_3$ solution. 
This is Weyl equivalent to $S^2 \times$ timelike-interval. The Weyl equivalence is acceptable because  $\partial \AdS_4$ is only defined within such Weyl rescaling. But to capture all of $\AdS_4^{\gl}$ we want $\CFT_3$ on all of $S^2 \times \mathbb{R}$, not just a time interval.

Fortunately if we switch to $\CGR_3$, then by its Weyl invariance, Weyl rescalings of  $\GR_3$ solutions are also solutions of $\CGR_3$~\cite{Horne:1988jf}. In particular  $S^2 \times \text{time-interval}$ must be a solution. By locality of $\CGR_3$ equations of motions, this means $S^2 \times \mathbb{R}$ is also a solution. We
can now couple $\CGR_3$ to $\CFT_3$ on $S^2 \times \mathbb{R}$. Given the CS form of $\CGR_3$, we expect states at fixed time $\tau = 0$ to transform under AS charges arising on the $S^2$ boundary at $\tau =0$ from $\CGR_3$ $=$ $SO(3,2)$ CS structure, and to persist in the  $\kappa \rightarrow \infty$ limit. This gives AS charges acting on  $\CFT_3$ states.
The full AS will 
contain the 
spacetime AS associated to $\CGR_3$ as well as any related to internal (CS) symmetries. The former has 
 $SO(3,2)$ global subalgebra. 
The SO(2) subgroup of SO(3,2) is just time translation in $\tau$, that is, the global $\AdS_4$/$\CFT_3$ Hamiltonian H. In particular all AS charges will have commutation relations with H, determining their $\tau$-dependence by the AS charge algebra.

\section{Mink$_4$ and Future Directions}
\label{sec:futureDirections}

In this paper, we generalized the notion of asymptotic symmetries (AS) applied to AdS$_4$, so that infinite-dimensional symmetries arise, analogous to the AS of Mink$_4$. We found a tight connection between these AS and the 3D holographic dual, in this case CFT$_3$, coupled to 3D gravity and Chern-Simons topological sectors. In 
turn, the combined 3D theory is dual to a CFT$_2$ structure in the sense of the AdS$_3$/CFT$_2$ correspondence, whose chiral currents and stress tensor house the AS charges. Several issues remain in order to fill out this story. 
Also, 
having seen these interconnections in AdS$_4$ quantum gravity and gauge theory, it is worth seeing if a parallel understanding can be gained for other 4D spacetimes where holography is less well understood, including Mink$_4$. 

\subsection{(A)dS$_4$}
It remains an important task to  explore how 3D gravity emerges from 
 $\AdS_4^{\Poincare}$  gravity as a (helicity-cut) soft or boundary limit  in analogy to 
 our discussion of $U(1)$ CS emerging from $\AdS^{\Poincare}_4$ $U(1)$ gauge theory. It will be interesting to see what type of 3D gravity emerges, GR$_3$ or CGR$_3$, or whether this depends on leading or subleading soft limits in some way. It will again be interesting to see if, and under what conditions, a finite effective level or central charge emerges. These  gauge and gravitational exercises should be repeated for 
  $\AdS_4^{\gl}$. Here we do not have the notion of soft limit since the spectrum is discrete, but the helicity-cut boundary limit continues to make sense. The connection between the emergent CS gauge fields and AS with 
   3D ``mirror" symmetry recast in dual 4D form \cite{Intriligator:1996ex,Kapustin:1999ha,Witten:2003ya} will be explored later~\cite{InPrep}. 
   
   We have explicitly given some infinite-dimensional subalgebras of the AS algebra acting on 
  $\AdS_4$ states,  while we have argued that the full AS algebra is implicitly captured by CGR$_3$ on $\partial$AdS$_4$.  It remains to explicitly describe this  algebra of AS charges acting on states of CFT$_3$ living on the $S^2$ boundary space.  
   
    By comparison with Mink$_4$, where IR divergences at loop level affect and complicate the soft limit \cite{Bern:2014oka,He:2014bga,Cachazo:2014dia,Bern:2014vva,He:2017fsb}, it is possible that $\AdS_4$ curvature IR-regulates and simplifies the considerations.  This remains to be explored.

It appears feasible to do a similar analysis in dS$_4$ as done here for AdS$_4$, and thereby discover AS in that case. It would be interesting to compare this approach to that of Ref~\cite{Hamada:2017gdg}. The approach suggested here would be compatible with the Poincare patch of dS.

 \subsection{AS as ``hair''}
 We have argued that infinite-dimensional AdS$_4$ AS are a useful form of ``hair'' for 4D black holes and other complex states, very much in the manner that the infinite-dimensional symmetry charges of
 CFT$_2$ characterize 2D states. Given how explicit this is in AdS$_4$, it would be interesting to explore whether the AS {\it fully characterize} any AdS$_4$ quantum state. Even a partial but still rich characterization may be relevant to the information puzzles of quantum black holes. The fact that AdS$_4$ gives a new 4D example of how soft fields take a 3D CS topological form, suggests that this phenomenon is more general, and should be understood on less symmetric (black hole) 4D spacetimes.

\subsection{$\Mink_4$}

In $\Mink_4$, $\text{Vir} \times \overline{\text{Vir}}$ super-rotations from the subleading soft limit of $\GR_4$ were shown to be captured by SO(3,1) CS = $\GR_3$ on $\text{(EA)dS}_3$~\cite{Cheung:2016iub}. But this CS description does not capture super-translations and the leading soft limit of GR$_4$, or even just Minkowski translations. It remains to find the 3D representation of all the soft limiting fields underlying the full $\XBMS_4$ AS. Doing so would be the analog of finding $\CGR_3$ = SO(3,2) CS for $\AdS_4$.

The most obvious guess would be to try CS gauging of the 4D Poincare group, $ISO(3,1)$. But there is a simple no-go  argument for this approach, in that there is no quadratic invariant to define the CS trace. The analogous $\GR_3$ on $\Mink_3$ is given by $ISO(2,1)$ CS, where the quadratic invariant is given by $\epsilon^{\mu \nu \rho} J_{\mu \nu} P_{\rho}$~\cite{Witten:1988hc}, obviously lacking 4D generalization.  

Nevertheless it is possible that a non-CS 3D characterization of $\Mink_4$ soft fields exists, reducing to $SO(3,1)$ CS for the subleading soft limit and super-rotations. 
The fact that the 2D conserved current housing super-translation KM charges in $\Mink_4$ was found to be a ECFT$_2$ descendent  operator of a partially conserved operator~\cite{Cheung:2016iub} suggests a role for {\it partially massless} gauge fields~\cite{Deser:1983tm,Deser:1983mm} in 3D, in turn
coupled to {\it partially conserved} currents of a 3D holographic dual of Mink$_4$ QG. 

One strategy to find this 3D characterization begins with the recently considered case of $\CGR_4$ on $\Mink_4$~\cite{Haco:2017ekf}. Here we may guess that the soft fields are characterized by SO(4,2) CS on $\text{(EA)dS}_3$, which does have the requisite quadratic invariant $J_{\Phi \Omega} J^{\Phi \Omega}$, $\Phi, \Omega = 0,1,2,3,4,5$.
This suggests that the 3D $SO(4,2)$ CS theory might be truncated (``Higgsed") to the 3D characterization of just 4D Poincare symmetric soft fields in terms of massless and partially massless 3D fields.
Another strategy is to see if there is a ``contraction" procedure for the $SO(3,2)$ CS description of AdS$_4$ AS found here that yields the 3D description of Mink$_4$ AS, in rough analogy to the contraction of $SO(2,2)$ CS governing AS of AdS$_3$ to the $ISO(2,1)$ CS governing AS of Mink$_3$.

A full 3D characterization of the soft Mink$_4$ fields would strongly constrain the form of a 
 3D holographic dual of 4D Mink QG, since the latter would have to be able to be coupled to the soft fields. This is in analogy to the neat compatibility of CFT$_3$  with coupling to CGR$_3$ in the AdS$_4$ context. 
 One can view such a connection in Mink$_4$ as a modern extension of Weinberg's classic derivation of consistency conditions on the S-matrix involving massless spin-$1$ and spin-$2$ particles. He showed~\cite{Weinberg:1964abc,Weinberg:1964ew,Weinberg:1965rz} by studying soft limits that matter necessarily has to couple to soft spin-$1$ through conserved charges and to soft spin-$2$ through gravitational-form charges satisfying the Equivalence Principle. But the full soft field structure may in fact be strong enough to prescribe the full holographic grammar of the dynamics. Such a grammar would effectively have to force the precise vanishing of the 4D cosmological constant, perhaps in a novel way.

\acknowledgments
RS would like to thank Clifford Cheung and Anton de la Fuente for earlier collaboration, insights and discussions related to this paper. In addition, the authors are grateful to Hamid Afshar, Nima Arkani-Hamed, Christopher Brust, Jared Kaplan, Juan Maldacena, Arif Mohd, Massimo Porrati and John Terning for helpful discussions and correspondence. This research was supported in part by the NSF under Grant No. PHY-1620074 and by the Maryland Center for Fundamental Physics (MCFP).

\bibliographystyle{JHEP}
\bibliography{References}

\end{document}